\documentclass[a4paper,twocolumn,superscriptaddress,11pt]{quantumarticlemod}
\pdfoutput=1
\usepackage[latin1]{inputenc}
\usepackage{amsmath,amsfonts,amssymb,amsthm,braket,graphicx,enumitem,framed,color,tikz,float,changes,mdframed,xpatch}
\usepackage{csquotes}
\usepackage{mdframed,xpatch}
\usepackage[vcentermath]{youngtab}
\usepackage[normalem]{ulem}
\theoremstyle{definition}
\makeatletter
   \xpatchcmd{\@thm}{\fontseries\mddefault\upshape}{}{}{} 
\makeatother
\newtheorem*{definition}{Definition}
\newtheorem{lemma}{Lemma}
\newtheorem*{theorem}{Theorem}
\newtheorem{thm}{Theorem}
\newtheorem*{postulate}{Postulate}
\newtheorem*{principle}{Principle}
\newtheorem*{opassum}{Operational Assumption}
\newtheorem*{opimp}{Operational Implication}

\newtheorem*{example}{Example}

\newtheorem*{test*}{Test}
\def\pcd{\mathbb{C}^d}

\def\pcthree{\mathbb{C}^3}

\def\pcnine{\mathbb{C}^9}
\def\pcdAdB{\mathbb{C}^{d_{\sf A}d_{\sf B}}}
\def\pcdA{\mathbb{C}^{d_{\sf A}}}
\def\pcdB{\mathbb{C}^{d_{\sf B}}}
\def\sud{\mathrm{SU}(d)}
\def\suthree{\mathrm{SU}(3)}

\def\sunine{\mathrm{SU}(9)}

\def\sudA{\mathrm{SU}(d_{\sf A})}
\def\sudB{\mathrm{SU}(d_{\sf B})}

\def\unity{\mathbb{I}}
\def\sA{{\sf A}}
\def\sB{{\sf B}}
\def\sC{{\sf C}}
\def\sl{{\rm L}}
\def\sg{{\rm G}}
\def\C{{\mathbb C}}
\def\kda{K_{d_\sA}}
\def\kdb{K_{d_\sB}}
\def\D{\mathcal D}
\def\F{\mathcal F}
\def\R{\mathbb R}
\newcommand{\ketbra}[2] {
	| #1 \rangle \! \langle #2 |}

\usepackage[numbers,sort&compress]{natbib}

\begin{document}
\title{Any modification of the Born rule leads to a violation of the purification and local tomography principles}
\date{\today}
\author{Thomas D. Galley}
\email{thomas.galley.14@ucl.ac.uk}
\affiliation{Department of Physics and Astronomy, University College London,
	Gower Street, London WC1E 6BT, United Kingdom}
\orcid{0000-0002-8870-3215}
\author{Lluis Masanes}
\orcid{0000-0002-1476-2327}
\affiliation{Department of Physics and Astronomy, University College London,
	Gower Street, London WC1E 6BT, United Kingdom}

\begin{abstract}
Using the existing classification of all alternatives to the measurement postulates of quantum theory we study the properties of bi-partite systems in these alternative theories. We prove that in all these theories the purification principle is violated, meaning that some mixed states are not the reduction of a pure state in a larger system. This allows us to derive the measurement postulates of quantum theory from the structure of pure states and reversible dynamics, and the requirement that the purification principle holds. The violation of the purification principle implies that there is some irreducible classicality in these theories, which appears like an important clue for the problem of deriving the Born rule within the many-worlds interpretation.
We also prove that in all such modifications the task of state tomography with local measurements is impossible, and present a simple toy theory displaying all these  exotic non-quantum phenomena. This toy model shows that, contrarily to previous claims, it is possible to modify the Born rule without violating the no-signalling principle.
Finally, we argue that the quantum measurement postulates are the most non-classical amongst all alternatives.
\end{abstract}

\maketitle

\section{Introduction}

\noindent
The postulates of quantum theory describe the evolution of physical systems by  distinguishing between the cases where observation happens or not. However, these postulates do not specify what constitutes observation, and it seems that an act of observation by one agent can be described as unperturbed dynamics by another~\cite{Everett_relative_1957}.
This opens the possibility of deriving the physics of observation within the picture of an agent-free universe that evolves unitarily. 
This problem has been studied within the dynamical description of quantum measurements~\cite{Allahverdyan_understanding_2013}, the decoherence program~\cite{Zurek_probabilities_2005} and the Many-Worlds Interpretation of quantum theory~\cite{Deutsch_quantum_1999,  Wallace_how_2010}.

In this work, instead of presenting another derivation of the measurement postulates, we take a more neutral approach and analyze all consistent alternatives to the measurement postulates. 
In particular, we prove that in each such alternative there are mixed states which are not the reduction of a pure state on a larger system. 
This property singles out the (standard) quantum measurement postulates including the Born Rule.

In our previous work~\cite{Galley_classification_2017} we constructed a complete classification of all alternative measurement postulates, by establishing a correspondence between these and certain representations of the unitary group.
However, this classification did not involve the consistency constraints that arise from the compositional structure of the theory; which governs how systems combine to form multi-partite systems.
In this work we take into account compositional structure, and prove that all alternative measurement postulates violate two compositional principles: purification~\cite{Chiribella_probabilistic_2010, Chiribella_informational_2011} and local tomography~\cite{Hardy_quantum_2001,Barrett_information_2005}.
We also present a simple alternative measurement postulate (a toy theory) which illustrates these exotic phenomena.
Additionally, this toy theory provides an interesting response to the claims that the Born rule is the only probability assignment consistent with no-signalling~\cite{Aaronson_quantum_2004, bao_grover_2016, Han_Quantum_2016}. 

In Section~\ref{Setup} we introduce a theory-independent formalism, which allows to study all alternatives to the measurement postulates. We also review the results of our previous work~\cite{Galley_classification_2017}.
In Section~\ref{features} we define the purification and local tomography principles, and show that these are violated by all alternative measurement postulates.
In Section~\ref{Toymodel} we describe a particular and very simple alternative measurement postulate, which illustrates our general results. 
In Section~\ref{Discussion} we discuss our results in the light of existing work. All proofs are in the appendices.

\section{Dynamically-quantum theories}\label{Setup}

\noindent
In this work we consider all theories that have the same pure states, dynamics and system-composition rule as quantum theory, but have a different structure of measurements and a different rule for assigning probabilities.

\subsection{States, transformations and composition postulates}

\noindent
The family of theories under consideration satisfy the following postulates, taken from the standard formulation of quantum theory.

\begin{postulate}[Quantum States]
Every finite-dimensional Hilbert space $\mathbb C^d$ corresponds to a type of system with pure states being the rays $\psi$ of $\mathbb C^d$.
\end{postulate}
  
\begin{postulate}[Quantum Transformations] 
The reversible transformations on the pure states $\mathbb C^d$ are $\psi \mapsto U\psi$ for all $U\in \sud$.
\end{postulate}
  
\begin{postulate}[Quantum Composition]
The joint pure states of systems $\mathbb C^{d_\sA}$ and $\mathbb C^{d_\sB}$ are the rays of $\mathbb C^{d_\sA} \otimes \mathbb C^{d_\sB} \simeq \mathbb C^{d_\sA d_\sB}$. 
\end{postulate}
  
\subsection{Measurement postulates}  
  
\noindent 
Before presenting the generalized  measurement postulate we need to introduce the notion of outcome probability function, or OPF. 
For each measurement outcome $x$ of system $\mathbb C^{d}$ there is a function $F^{(x)}$ that assigns to each ray $\psi$ in $\mathbb C^d$ the probability $F^{(x)} (\psi) \in [0,1]$ for the occurrence of outcome $x$.
Any such function $F$ is called an OPF.
Each system has a (trivial) measurement with only one outcome, which must have probability one for all states. The uniqueness of this trivial measurement is the Causality Axiom of~\cite{Chiribella_probabilistic_2010}.  The OPF associated to this outcome is called the unit OPF $\bf u$, satisfying ${\bf u}(\psi)=1$ for all $\psi$. 
A $k$-outcome measurement is a list of $k$ OPFs $(F^{(1)}, \ldots, F^{(k)})$ satisfying the normalization condition $\sum_i F^{(i)} =\bf u$.  It is not necessarily the case that every list of OPFs satisfying this condition defines a measurement, though this assumption can be made.
As an example, the OPFs of quantum theory are the functions 
\begin{equation}
  \label{Q OPFs}
  F(\psi) = 
  {\rm tr} (\hat F \ketbra{\psi}{\psi})\ ,
\end{equation}
for all Hermitian matrices $\hat F$ satisfying $0\leq \hat F \leq \unity$. 
This implies that $\hat {\bf u} = \unity$.
(Here and in the rest of the paper we assume that kets $\ket \psi$ are normalized.)

\begin{postulate}[Alternative Measurements] 
Every type of system $\mathbb C^d$ has a set of OPFs $\mathcal F_d$ with a bilinear associative product $\star : \mathcal F_{d_\sA} \times \mathcal F_{d_\sB} \to \mathcal F_{d_\sA d_\sB}$ satisfying the following consistency constraints:
\begin{itemize}
  \item[\bf C1.] For every $F\in \mathcal F_d$ and $U\in \sud$ there is an $F' \in \mathcal F_d$ such that $F'(\psi) = F(U\psi)$ for all $\psi\in \mathbb C^d$.
  That is, the composition of a unitary and a measurement can be globally considered a measurement.

  \item[\bf C2.] For any pair of different rays $\psi \neq \phi$ in $\mathbb C^d$ there is an $F\in \mathcal F_d$ such that $F(\psi) \neq F(\phi)$.
  That is, different pure states must be operationally distinguishable.
    
  \item[\bf C3.] The $\star$-product satisfies $\bf u_\sA \star u_\sB = u_{\sA \sB}$ and 
\begin{equation}\label{starprod}
  (F_\sA \star F_\sB) 
  (\psi_\sA \otimes \phi_\sB)
  = F_\sA (\psi_\sA) F_\sB (\phi_\sB)\ ,
\end{equation}
for all $F_\sA \in \mathcal F_{d_\sA}$, $F_\sB \in \mathcal F_{d_\sB}$, $\psi_\sA \in \mathbb C^{d_\sA}$, $\phi_\sB \in \mathbb C^{d_\sB}$.    
That is, tensor-product states $\psi_\sA \otimes \phi_\sB$ contain no correlations.

  \item[\bf C4.] For each $\phi_{\sA \sB} \in \mathbb C^{d_\sA} \otimes \mathbb C^{d_\sB}$ and $F_\sB \in \mathcal F_{d_\sB}$ there is an ensemble $\{(\psi^i_\sA, p_i)\}_i$ in $\mathbb C^{d_\sA}$ such that
\begin{equation}
  \frac
  {(F_\sA \star F_\sB) 
  (\phi_{\sA\sB})}
  {({\bf u}_\sA \star F_\sB) 
  (\phi_{\sA\sB})}
  = \sum_i p_i
  F_\sA (\psi^i_\sA)
  \ ,
\end{equation}
for all $F_\sA \in \mathcal F_{d_\sA}$.
That is, the reduced state on $\sA$ conditioned on outcome $F_\sB$ on $\sB$ (and re-normalized) is a valid mixed state of $\sA$. In the next sub-section we fully articulate the notions of ensemble and mixed state.    
    
  \item[\bf C5.] Consider measurements on system $\mathbb C^{d_\sA}$ with the help of an ancilla $\mathbb C^{d_\sB}$. For any ancillary state $\phi_\sB \in \mathbb C^{d_\sB}$ and any OPF in the composite $F_{\sA \sB} \in \mathcal F_{d_\sA d_\sB}$ there exists an OPF on the system $F'_\sA \in \mathcal F_{d_\sA}$ such that
  \begin{equation}
    F'_\sA (\psi_\sA)
    =
    F_{\sA \sB} (\psi_\sA \otimes \phi_\sB)
  \end{equation}
for all $\psi_\sA$.

\end{itemize}
\end{postulate}

\noindent
The derivation of these consistency constraints from operational principles is provided in Appendix \ref{OPFCons}.
Continuing with the example of quantum theory~\eqref{Q OPFs}, the $\star$-product in this case is
\begin{equation}
  \label{compo Q}
  (F_\sA \star F_\sB) (\psi_{\sA \sB})
  =
  {\rm tr} (\hat F_\sA \otimes \hat F_\sB \ketbra{\psi_{\sA \sB}}{\psi_{\sA \sB}})
  \ .
\end{equation}
A trivial modification of the Measurement Postulate consists of taking that of quantum mechanics~\eqref{Q OPFs} and restricting the set of OPFs in some way, such that not all POVM elements $\hat F$ are allowed.
In this work, when we refer to ``all alternative measurement postulates" we do not include these trivial modifications.

\subsection{Mixed states and the Finiteness Principle} 

\noindent
A source of systems that prepares state $\psi_i \in \mathbb C^d$ with probability $p_i$ is said to prepare the ensemble $\{(\psi_i, p_i)\}_i$.
Two ensembles $\{(\psi_i, p_i)\}_i$ and $\{(\phi_j, q_j)\}_j$ are equivalent if they are indistinguishable
\begin{equation}
  \sum_i p_i F(\psi_i)
  =
  \sum_j q_j F(\phi_j)
\end{equation}
for all measurements $F\in \mathcal F_d$.
Note that distinguishability is relative to the postulated set of OPFs $\mathcal F_d$.
A mixed state $\omega$ is an equivalence class of indistinguishable ensembles, and hence, the structure of mixed states is also relative to $\mathcal F_d$. To evaluate an OPF $F$ on a mixed state $\omega$ we can take any ensemble $\{(\psi_i, p_i)\}_i$ of the equivalence class $\omega$ and compute
\begin{equation}
  F(\omega) = \sum_i p_i F(\psi_i)\ .
\end{equation}
In general, ensembles can have infinitely-many terms, hence, the number of parameters that are needed to characterize a mixed state can be infinite too. When this is the case, state estimation without additional assumptions is impossible, and for this reason we make the following assumption.

\begin{principle}[Finiteness]
Each mixed state of a finite-dimensional system $(\mathbb C^d, \mathcal F_d)$ can be characterized by a finite number $K_d$ of parameters. 
\end{principle}

\noindent
Recall that in quantum theory we have $K_d = d^2-1$. And in general, the distinguishability of all rays in $\mathbb C^d$ implies $K_d \geq 2 d-2$. %
These $K_d$ parameters can be chosen to be a fix set of ``fiducial" OPFs $F_1, \ldots, F_{K_d} \in \mathcal F_d$, which can be used to represent any mixed state $\omega$ as
\begin{equation}
  \bar \omega = \left(
  \begin{array}{c}
    F_1 (\omega) \\
    F_2 (\omega) \\
    \vdots \\
    F_{K_d} (\omega)
  \end{array} \right) \ .
\end{equation}
The fact that OPFs are probabilities implies that any OPF $F\in \mathcal F_d$ is a linear function of the fiducial OPFs
\begin{equation}
  \label{basis}
  F = \sum_i c_i F_i\ .
\end{equation}
In other words, the fiducial OPFs $F_1, \ldots, F_{K_d} \in \mathcal F_d$ constitute a basis of the real vector space spanned by $\mathcal F_d$.
Using the consistency constraint {\bf C1}, we define the $\sud$ action 
\begin{equation}\label{sudaction}
  F_i = 
  \sum_{i'} \bar \Gamma_{i,i'} (U)\, F_{i'}
\end{equation}
on the vector space spanned by $\mathcal F_d$.
This associates to system $(\mathbb C^d, \mathcal F_d)$ a $K_d$-dimensional representation of the group $\sud$.
This, together with the other consistency constraints, implies that only certain values of $K_d$ are allowed. For example $K_2 = 3, 7, 8, 10, 11, 12, 14\ldots$

\subsection{Measurement postulates for single systems}

\noindent
In this subsection we review some of the results obtained in~\cite {Galley_classification_2017}. These provide the complete classification of all sets $\mathcal F_d$ satisfying the Finiteness Principle and the consistency constraints {\bf C1} and {\bf C2}. 
These results ignore the existence of the $\star$-product, {\bf C3}, {\bf C4} and {\bf C5}. Hence, in alternative measurement postulates with a consistent compositional structure there will be additional restrictions on the valid sets $\mathcal F_d$. This is studied in Section~\ref{features}.

\begin{theorem}[Characterization]\label{Caracterisation}
If $\mathcal F_d$ satisfies the Finiteness Principle and {\bf C1} then there is a positive integer $n$ and a map $F \mapsto \hat F$ from $\mathcal F_d$ to the set of $d^n \times d^n$ Hermitian matrices such that
\begin{equation}
    F(\psi) = {\rm tr}\!\left( \hat F
    |\psi\rangle\!\langle \psi|^{\otimes n}
    \right)
\end{equation}
for all normalized vectors $\psi \in \mathbb C^d$.
\end{theorem}

\noindent
Note that there are many different sets $\mathcal F_d$ with the same $n$. In particular, since the $\sud$ action 
\begin{equation}
  \label{red rep}
  |\psi\rangle\!\langle \psi|
  ^{\otimes n} 
  \mapsto 
  U^{\otimes n} 
  |\psi\rangle\!\langle \psi|
  ^{\otimes n} 
  U^{\otimes n \dagger}
\end{equation}
is reducible, the Hermitian matrices $\hat F$ can have support on the different irreducible sub-representations of~\eqref{red rep}, generating sets $\mathcal F_d$ with very different physical properties.
All these possibilities are analyzed in~\cite{Galley_classification_2017}. 

\begin{theorem}[Faithfulness]
If $\mathcal F_d$ satisfies the Finiteness Principle, {\bf C1} and {\bf C2}, then
\vspace{-1mm}
\begin{description}
\setlength{\itemsep}{0pt}
\item[\bf case $d\geq 3$] there is a non-constant $F\in \mathcal F_d$.

\item[\bf case $d=2$] there is $F\in \mathcal F_2$ such that $\hat F$ has support on a sub-representation of the $\mathrm{SU}(2)$ action~\eqref{red rep} with odd angular momentum.

\end{description}
\end{theorem}

\section{Features of all alternative measurement postulates}\label{features}

\noindent
In this section we analyze the compositional structure of alternative measurement postulates. We do so by considering two well known physical principles which, together with other assumptions, have been used to reconstruct the full formalism of quantum theory~\cite{Masanes_derivation_2011,Chiribella_informational_2011,Dakic_quantum_2011,Barnum_local_2014}.
Remarkably, these principles are
violated by all alternative measurement postulates.

\subsection{The Purification Principle}

\noindent
This principle establishes that any mixed state is the reduction of a pure  state in a larger system.
This legitimises the ``Church of the Larger Hilbert Space", an approach to physics that always assumes a global pure state when an environment is added to the systems under consideration~\cite{Nielsen_quantum_2011}.

\begin{principle}[Purification]
For each ensemble $\{(\psi_i, p_i)\}_i$ in $\mathbb C^{d_\sA}$ there exists a pure state $\phi_{\sA \sB}$ in $\mathbb C^{d_\sA} \otimes \mathbb C^{d_\sB}$ for some $d_\sB$ satisfying
\begin{equation}
  (F_{\sA} \star {\bf u}_{\sB})(\phi_{\sA \sB})
  = 
  \sum_i p_i F_{\sA} (\psi_i) \ ,
\end{equation}
for all $F_{\sA} \in \mathcal F_{d_\sA}$.
\end{principle}

\noindent 
Note that the original version of the purification principle introduced in~\cite{Chiribella_probabilistic_2010} additionally demands that the purification state $\phi_{\sA \sB}$ is unique up to a unitary transformation on $\mathbb C^{d_\sB}$.
Also note that the following theorem does not require the Finiteness Principle.

\begin{theorem}[No Purification]
All alternative measurements postulates $\mathcal F_d$ satisfying {\bf C1}, {\bf C2}, {\bf C3} and {\bf C4} violate the purification principle.
\end{theorem}

\noindent
This implies that in all alternative measurement postulates there are operational processes, such as mixing two states, which cannot be understood as a reversible transformation on a larger system. 
In such alternative theories, agents that perform physical operations cannot be integrated in an agent-free universe, as can be done in quantum theory, and the Church of the Larger Hilbert Space is illegitimate. 
The assumption that agents'	 actions can be understood as reversible transformations on a larger system is also the starting point of the many-worlds interpretation of quantum theory.
Hence, it seems like the no-purification theorem provides important clues for the derivation of the Born Rule within the many-worlds interpretation~\cite{Everett_relative_1957,Zurek_probabilities_2005,Wallace_how_2010,Deutsch_quantum_1999}. In section~\ref{Zurek} we discuss further the possibility of using this result to derive the quantum measurement postulates, and contrast it to Zurek's envariance based derivation of the Born rule.

\subsection{The Local Tomography Principle}

\noindent
This principle has been widely used in reconstructions of quantum theory and the formulation of alternative toy theories~\cite{Dakic_quantum_2011, Masanes_derivation_2011, Chiribella_informational_2011,Barnum_local_2014, Masanes_existence_2013, Hohn_toolbox_2017}. One of the reasons is that it endows the set of mixed states with a tensor-product structure~\cite{Barrett_information_2005}.
This principle states that any bi-partite state is characterized by the correlations between local measurements.
That is, two different mixed states $\omega_{\sA\sB} \neq \omega'_{\sA\sB}$ on $\mathbb C^{d_\sA} \otimes \mathbb C^{d_\sB}$ must provide different outcome probabilities
\begin{equation}
  (F_{\sA} \star F_{\sB})(\omega_{\sA\sB}) 
  \neq
  (F_{\sA} \star F_{\sB})(\omega'_{\sA\sB})
\end{equation}
for some local measurements $F_{\sA} \in \mathcal F_{d_\sA}$, $F_{\sB} \in \mathcal F_{d_\sB}$.
Using the notation introduced in~\eqref{basis} we can formulate this principle as follows.

\begin{principle}[Local Tomography]
If $\{F_a \}_a$ is a basis of $\mathcal F_{d_\sA}$ and $\{F_b \}_b$ is a basis of $\mathcal F_{d_\sB}$ then $\{F_a \star F_b \}_{a,b}$ is a basis of $\mathcal F_{d_\sA d_\sB}$, where $a=1, \ldots, K_{d_\sA}$ and $b=1, \ldots, K_{d_\sB}$. 
\end{principle}

\noindent
A theory is said to violate local tomography if at least one composite system within the theory violates local tomography.
Therefore, it is sufficient to analyze the particular bi-partite system $\mathbb C^3 \otimes \mathbb C^3 = \mathbb C^9$.

\begin{theorem}[No Local Tomography]
All alternative measurements $\mathcal F_3$ and $\mathcal F_9$ satisfying {\bf C1}, {\bf C2}, {\bf C3} and the Finiteness Principle violate the local tomography principle.
\end{theorem}
\noindent
The above result is proven in Appendix \ref{nolocaltom}. We first show that all transitive theories which obey the local tomography principle have a group action acting on the mixed states which has a certain structure. Any group action of the form \eqref{sudaction} which does not have this structure must correspond to a system which violates local tomography. We show that all representations of $\sunine$ which correspond to systems with alternative measurement postulates do not have this structure. This entails that all non-quantum $\C^9$ systems which are composites of two $\C^3$ systems violate local tomography.

The technical result proven to show this is the following. All non-quantum irreducible representations of $\sunine$ which are sub-representations of the action~\eqref{red rep} have a sub-representation $1^{3} \otimes 1^{3}$ when restricted to the subgroup $\suthree \times \suthree$. Here $1^3$ denotes the trivial representation of $\suthree$.

\subsubsection{Comment on $\mathbb{C}^2 \otimes \mathbb{C}^2$ systems}

In quantum theory any  $\C^d$ ($d \geq 2$) system can be simulated using some number of qubits. In this sense qubits can be viewed as fundemental information units~\cite{Masanes_existence_2013}. Since $\mathbb C^2$ systems have a priveleged status it is natural to ask whether $\mathbb{C}^4 = \mathbb{C}^2 \otimes \mathbb{C}^2$ systems in theories with modified measurement postulates are locally tomographic. The proof technique used for the No Local Tomography Theorem applies only to a certain family of these $\mathbb{C}^4$ systems. However all instances of $\mathbb{C}^4$ systems studied by the authors which were not part of this family were found to not be locally tomographic. We conjecture that all $\mathbb{C}^4 = \mathbb{C}^2 \otimes \mathbb{C}^2$ violate the local tomography principle.

\section{A Toy Theory}\label{Toymodel}

\noindent
In this section we present a simple alternative measurement postulate $(\mathcal F_d, \star)$ which serves as example for the results that we have proven in general (violation of the purification and local tomography Principles). 
In Appendix \ref{toymodel} it is proven that this alternative measurement postulate satisfies all consistency constraints ({\bf C1}, {\bf C2}, {\bf C3}, {\bf C4}, {\bf C5}) except for the associativity of the $\star$-product. 
This implies that this toy theory is only fully consistent when dealing with single and bi-partite systems.
However, most of the work in the field of general probabilistic theories (GPTs) focuses on bi-partite systems, because these already display very rich phenomenology. In the following we consider two local subsystems of dimension $d_\sA$ and $d_\sB$  with sets of OPFs $\mathcal F_{d_\sA}^\sl$ and $\mathcal F_{d_\sB}^\sl$. The composite (global) system has a set of OPFs $\mathcal F_{d_\sA d_\sB}^\sg$.

\begin{definition}[Local effects $\mathcal F_d^\sl$]
Let $S$ be the projector onto the symmetric subspace of $\mathbb C^d \otimes \mathbb C^d$. To each $d^2 \times d^2$ Hermitian matrix $\hat F$ satisfying
\begin{itemize}
  \item $0\leq \hat F \leq S$,
  
  \item $\hat F = \sum_i \alpha_i \ketbra{\phi_i}{\phi_i}^{\otimes 2}$ for some $\ket{\phi_i} \in \mathbb C^d$ and $\alpha_i >0$, 
  
  \item $S-\hat F = \sum_i \beta_i \ketbra{\varphi_i}{\varphi_i}^{\otimes 2}$ for some $\ket{\varphi_i} \in \mathbb C^d$ and $\beta_i >0$, 
\end{itemize}
there corresponds the OPF 
\begin{equation}
  \label{toy F}
  F (\psi) = 
  {\rm tr} \left(\hat F \ketbra{\psi}{\psi}^{\otimes 2} \right)
  \ ,
\end{equation}
The unit OPF corresponds to $\hat {\bf u} =S$.
\end{definition}

\noindent
That is, both matrices, $\hat F$ and $S-\hat F$, have to be not-necessarily-normalized mixtures of symmetric product states.

\begin{example}[Canonical measurement for $d$ prime]
For the case where $d$ is prime there exists a canonical measurement which can be constructed as follows. Consider the $(d+1)$ mutually unbiased bases (MUBs): $\{ \ket{\phi_i^j}  \}_{i=1}^d$ where $j$ runs from $1$ to $d+1$~\cite{Bandyopadhyay_new_2001}. Then we can associate an OPF to each Hermitian matrix $\frac{1}{2} \ketbra{\phi_i^j}{\phi_i^j}^{\otimes 2}$. Since the basis elements of these MUBs form a complex projective 2-design~\cite{Klappenecker_mutually_2005}, by the definition of 2-design~\cite{Zhu_clifford_2016}, we have the normalization constraint:
\begin{equation}
\frac{1}{2} \sum_{i,j}  \ketbra{\phi_i^j}{\phi_i^j}^{\otimes 2} = S \ ,
\end{equation}
and hence the set of OPFs forms a measurement.
\end{example}

\begin{definition}[$\star$ product] For any pair of OPFs $F_{\sA} \in \mathcal F_{d_\sA}^\sl$ and $F_{\sB} \in \mathcal F_{d_\sB}^\sl$ the Hermitian matrix corresponding to their product $F_\sA \star F_\sB \in \mathcal F_{d_\sA d_\sB}^\sg$ is
\begin{equation}
  \label{toy * product}
  \widehat{F_\sA \star F_\sB} = 
  \hat F_\sA \otimes \hat F_\sB + 
  \frac {{\rm tr}\hat F_\sA}{{\rm tr} S_\sA} A_\sA \otimes   
  \frac {{\rm tr}\hat F_\sB}{{\rm tr} S_\sB}  A_\sB 
  \ ,
\end{equation}
where $S_\sA$ and $A_\sA$ are the projectors onto the symmetric and anti-symmetric subspaces of $\mathbb C^{d_\sA} \otimes \mathbb C^{d_\sA}$, and analogously for $S_\sB$ and $A_\sB$. 
\end{definition}

\noindent
This product is clearly bilinear and, by using the identity $S_{\sA\sB} = S_\sA \otimes S_\sB + A_\sA \otimes A_\sB$, we can check that ${\bf u}_\sA \star {\bf u}_\sA = {\bf u}_{\sA\sB}$.

We observe that not all effects $\widehat{F_\sA \star F_\sB}$ are of the form $\sum_i \alpha_i \ketbra{\phi_i}{\phi_i}_{\sA \sB}^{\otimes 2}$. Hence the set of effects on the joint system is not $\mathcal F_{d_\sA d_\sB}^\sl$, but has to be extended to $\mathcal F_{d_\sA d_\sB}^\sg$ to include these joint product effects.

\begin{definition}[Global effects $\mathcal F_{d_\sA d_\sB}^\sg$]
The set $\mathcal F_{d_\sA d_\sB}^\sg$ should include all product OPFs $\widehat{F_\sA \star F_\sB}$, all OPFs $\mathcal F_{d_\sA d_\sB}^\sl$ of $\mathbb C^{d_\sA d_\sB}$ understood as a single system, and their convex combinations.
\end{definition}

The identity $S_{\sA\sB} = S_\sA \otimes S_\sB + A_\sA \otimes A_\sB$ perfectly shows that the vector space $\mathcal F_{d_\sA d_\sB}^\sg$ is larger than the tensor product of the vector spaces $\mathcal F_{d_\sA}^\sl$ and $\mathcal F_{d_\sB}^\sl$, by the extra term $A_\sA \otimes A_\sB$.
This implies that this toy theory violates the Local-Tomography Principle.

\begin{figure}
	\centering{
\includegraphics[width=0.3\textwidth]{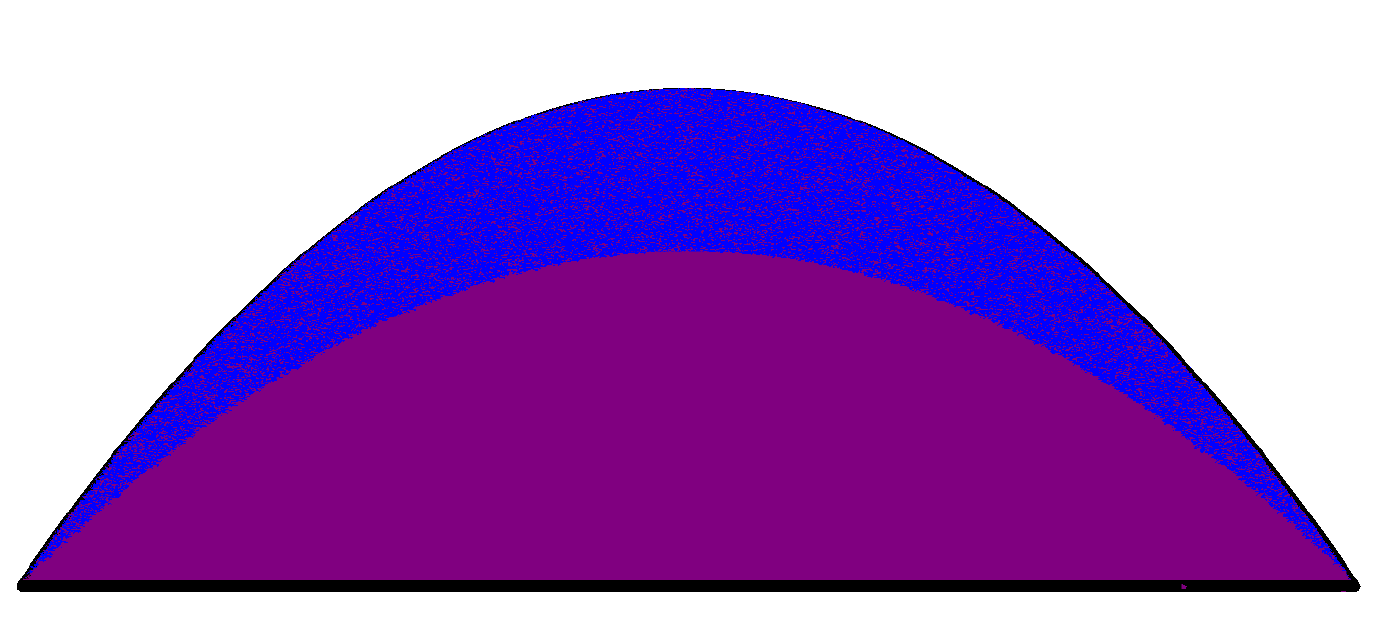}
}
\caption{Projection of $\mathbb C^2$ toy model state space. In this projection the coloured points (blue and purple) are states of the form $\sum_i p_i \ketbra{\psi_i}{\psi_i}^{\otimes 2}$. The blue points are projections of reduced states of a larger system obtained using formula \eqref{reduced state}. The left hand corner corresponds to the state $\ketbra{0}{0}^{\otimes 2}$ and the right hand corner to the state $\ketbra{1}{1}^{\otimes 2}$. All pure states are projected onto the curved boundary. This figure shows that there are local mixed states (in purple) which are not reduced states (in blue).}
\label{purificationpic}
\end{figure}

The joint probability of outcomes $F_\sA$ and $F_\sB$ on the entangled state $\psi_{\sA \sB} \in \mathbb C^{d_\sA} \otimes \mathbb C^{d_\sB}$ can be written as
\begin{align*}
  & (F_\sA \star F_\sB) (\psi_{\sA \sB})  \\
= 	&  {\rm tr}\! \left[ \left(\hat F_\sA \otimes \hat F_\sB +   \mbox{$\frac {{\rm tr}\hat F_\sA}{{\rm tr} S_\sA}$}  A_\sA \otimes 
 \mbox{$\frac {{\rm tr}\hat F_\sB}{{\rm tr} S_\sB}$} A_\sB \right) 
 \!\ketbra{\psi_{\sA\sB}}{\psi_{\sA\sB}}^{\otimes 2} \right] \ .
\end{align*}
When we only consider sub-system $\sA$ outcome probabilities are given by
{\small
\begin{eqnarray}
  \nonumber
  & & (F_\sA \star {\bf u}_\sB) (\psi_{\sA \sB})
  \\ &=&
  {\rm tr}\! \left[ \left(\hat F_\sA \otimes S_\sB +   \mbox{$\frac {{\rm tr}\hat F_\sA}{{\rm tr} S_\sA}$}  A_\sA \otimes 
 A_\sB \right) \ketbra{\psi_{\sA\sB}}{\psi_{\sA\sB}}^{\otimes 2} \right]
  \\ &=&
  {\rm tr}_\sA\! \left[ \hat F_\sA\, \bar \omega_\sA \right] \ ,
\end{eqnarray}
}
where the reduced state must necessarily be
{\small
\begin{equation}
  \label{reduced state}
  \bar \omega_\sA 
  =
  {\rm tr}_\sB \! \left( S_\sB \ketbra{\psi_{\sA\sB}}{\psi_{\sA\sB}}^{\otimes 2} \right)
  + \frac {S_\sA}{{\rm tr} S_\sA} 
  {\rm tr} \! 
  \left(A_\sA  A_\sB 
  \ketbra{\psi_{\sA\sB}}{\psi_{\sA\sB}}^{\otimes 2} \right)
\end{equation}
}
All these reductions $\bar\omega _\sA$ of pure bipartite states $\psi_{\sA\sB}$ are contained in the convex hull of $\ketbra{\phi_{\sA}}{\phi_{\sA}}^{\otimes 2}$, as required by the consistency constraint ${\bf C4}$.
However, not all mixtures of $\ketbra{\phi_{\sA}}{\phi_{\sA}}^{\otimes 2}$ can be written as one such  reduction~\eqref{reduced state}. That is, the purification postulate is violated. This phenomenon is graphically shown in Figure \ref{purificationpic}.

This toy model violates the \emph{no-restriction hypothesis}~\cite{Chiribella_probabilistic_2010}, in that not all mathematically allowed effects on the local state spaces are allowed effects. It also violates the principle of \emph{pure sharpness}~\cite{Chiribella_operational_2015} in that all the effects are noisy.

\section{Discussion}\label{Discussion}

\subsection{Interpreting results as a derivation of the Born rule}\label{Zurek}

In this paper we have shown that all modifications to the quantum measurements lead to violations of the purification and local tomography principles. This entails that one can derive the measurement postulates of quantum theory from the structure of pure states and dynamics  and either the assumption of local tomography or purification. Such a derivation uses the operational framework which can be viewed as a background assumption.

A derivation of the Born rule which starts from similar assumptions to ours, but not within an operational setting, is the envariance based derivation of Zurek~\cite{Zurek_probabilities_2005}. Zurek begins by assuming the dynamical structure of quantum theory and the assumption that quantum theory is \emph{universal}, which is to say that all the phenomena we observe can be explained in terms of quantum systems interacting. Specifically the classical worlds of devices can be modelled quantum mechanically, including the measurement process. We observe that this is philosophically very different to the operational approach adopted in this work, which takes the classical world as a primitive. By assuming the dynamical structure of quantum theory  and the assumption of universality (as well as some auxiliary assumptions) Zurek shows that measurements are associated to orthonormal bases, and that outcome probabilities are given by the Born rule. For criticisms of Zurek's approach we refer the reader to~\cite{Caves_note_2004, Schlosshauer_zureks_2005, Barnum_no_2003, mohrhoff_probabilities_2004}.

We observe that the purification postulate seems linked to the notion that quantum theory is universal, in the sense that any classical uncertainty can be explained as originating from some pure global quantum state. This shows an interesting link to Zurek's derivation, since although we work within an operational framework, the concept of purification is linked to the idea that quantum theory is universal. This shows that we can also rely on a concept linked to universality in order to derive the Born rule (and the structure of measurements) within an operational approach.

We observe that we can also derive the measurement postulates of quantum theory from the assumption of local tomography, which does not have this connotation of universality.

\subsection{No-signalling}

\noindent
Multiple proofs have been put forward which claim to show that violations of the Born rule lead to signalling \cite{Aaronson_quantum_2004, bao_grover_2016,Han_Quantum_2016}. However the authors only consider modifications of the Born rule of a specific type.
In \cite{Aaronson_quantum_2004, bao_grover_2016} the authors only consider modifications of the Born rule of the following form:
\begin{equation}\label{pnormrule}
p(k | \psi ) = \frac{| \! \braket{k|\psi} \! |^n}{\sum_{k'} | \! \braket{k'|\psi} \! |^n} \ ,
\end{equation}
where $\ket \psi = \sum_k \alpha_k \ket k$. Modifications of this form are very restricted. 
By modifying all the measurement postulates of quantum theory, we can create toy models like the one introduced, which are non-signalling (as we will show momentarily). This shows that by modifying the Born rule in a more general manner one can avoid issues of signalling.

In the case of the toy model it is immediate to see that it is consistent with no-signalling. The condition of no-signalling is equivalent to the existence of a well defined state-space for the subsystem (i.e. independent of action on the other subsystem). We see then that no-signalling is just a consequence of there existing a well defined reduced state~\cite{Chiribella_probabilistic_2010}. 

\subsection{Purification as a constraint on physical theories}

In Theorem 19 of~\cite{Chiribella_probabilistic_2010} the authors show that any two convex theories with the same states (pure and mixed) which obey purification are the same theory. In other words ``states specify the theory" for theories with purification~\cite{Chiribella_probabilistic_2010}. In this chapter  we show that in the case of theories with pure states ${\rm P}\C^d$ and dynamical group $\sud$, any two theories  which obey purification with the same pure states and reversible dynamics are the same. This means that for a restricted family of theories (those with systems with pure states ${\rm P}\C^d$ and dynamical group $\sud$) we have the same result as Theorem 19 of~\cite{Chiribella_probabilistic_2010} but with different assumptions. It would be interesting to establish whether this is a general feature of theories  with purification, namely that the pure states and reversible dynamics specify the theory.

\subsection{The quantum measurement postulates are the most non-classical}

One consequence of the classification in \cite{Galley_classification_2017} is that the quantum measurement postulates are the ones which give the lowest dimensional state spaces. In this sense the quantum measurement postulates are the most non-classical, since they give rise to the state spaces with the higest degree of indistinguishable ensemles~\cite{Mielnik_generalized_1974}. We remember that in classical probability theory all ensembles are distinguishable.

The violation of purification seems to indicate some other, distinct type of classicality. In theories which violate purification there are some preparations which can only be modelled as arising from a classical mixture of pure states. There appears to be some sort of irreducible classicality. However in theories which obey purification we can always model such preparations as arising from the reduction of a global pure state. Since the quantum measurement postulates are the only ones which give rise to systems which obey purification they can be viewed as the most non-classical amongst all alternatives. 

Hence we see that according to these two distinct notions of non-classicality the quantum measurement postulates are the most non-classical amonst all possible measurement postulates.

\subsection{Toy model}

The toy model can be obtained by restricting the states and measurements of two pairs of quantum systems $(\mathbb C^{d_\sA})^{\otimes 2}$ and $(\mathbb C^{d_\sB})^{\otimes 2}$. In this sense we obtain a theory which violates both local tomography and purification. This method of constructing theories is similar to real vector space quantum theory, which can also be obtained from a suitable restriction of quantum states and also violates local tomography. The main limitation of the toy model is that it does not straightforwardly extend to more than two systems. There is a natural generalisation of the toy model to consider effects to be linear in $\ketbra{\psi}{\psi}^{\otimes n}$ for $n >2$, however showing the consistency of the reduced state spaces and joint effects is more complex.

\subsection{Theories which decohere to quantum theory}

A recent result~\cite{Lee_nogo_2017} shows that all operational theories which decohere to quantum theory (in an analogous way to which quantum theory decoheres to classical theory) must violate either purification or causality (or both). The authors define a hyper-decoherence map which maps states of the post-quantum system to states of a quantum system (embedded within the post-quantum system). This map obeys the following properties:
\begin{enumerate}
	\item Terminality: applying the map followed by the unit effect is equivalent to just applying the unit effect.
	\item Idempotency: applying the map twice is the same as applying it once.
	\item The pure states of the quantum subsystem are pure states of the post-quantum system. Similarly the maximally mixed state of the quantum subsystem is the maximally mixed state of the post-quantum system. \label{purity}
\end{enumerate}

Now we ask whether the systems in this paper (which violate purification) can correspond to post-quantum systems which decohere to quantum systems in some reasonable manner. We consider the toy model with pure states $\ketbra{\psi}{\psi}^{\otimes 2}$ and no restriction on the allowed effects (this toy model is only valid for single systems). Now consider a linear map from $\ketbra{\psi}{\psi}^{\otimes 2}$ to some embedded quantum system such that the pure states of the system are also of the form $\ketbra{\phi}{\phi}^{\otimes 2}$ (by property \ref{purity}). This map must be of the form $U^{\otimes 2} \ketbra{\psi}{\psi}^{\otimes 2} U^{\dagger \otimes 2}$. Hence its action on the state space ${\rm conv} \left( \ketbra{\phi}{\phi}^{\otimes 2} \right)$ gives an identical state space and the map is trivial.\footnote{We thank referee 1 for this observation.} This shows that the only hyper-decoherence maps which obey all the conditions set out by Lee and Selby must be trivial.
 
As we argue later, it is not immediately obvious that a hyper-decoherence map should obey condition~\ref{purity}. We now define a hyper-decoherence like map which does not meet this condition.
Consider the following map where we label the copies of $\ketbra{\psi}{\psi}$, $1$ and $2$:
\begin{align}\label{decmap}
\D(\ketbra{\psi}{\psi}^{\otimes 2}) & = {\rm Tr}_2 (\ketbra{\psi}{\psi}^{\otimes 2}) \otimes \ketbra{0}{0}  \\
& = \ketbra{\psi}{\psi} \otimes \ketbra{0}{0} \ .
\end{align}

This meets properties 1 and 2, but not 3. Indeed the states $\ketbra{\psi}{\psi} \otimes \ketbra{0}{0}$ are not valid mixed states of the post-quantum system. The image of this map $\ketbra{\psi}{\psi} \otimes \ketbra{0}{0}$ for all $\psi \in {\rm P} \C^d$ is a quantum state space.

The authors of \cite{Lee_nogo_2017} show that requirement 3 can be replaced by the requirement that the hyper-decoherence map maps between systems with the same information dimension. As shown in~\cite{Galley_classification_2017} (for the case ${\rm P} \C^2$) the information dimension of an unrestricted $\ketbra{\psi}{\psi}^{\otimes 2}$ state space is larger than that of a qubit. 

The hyper-decoherence map of \cite{Lee_nogo_2017} is inspired by the decoherence map which exists between quantum and classical state spaces. The map introduced in this section is very different from this, since its image is not an embedded state space; however it may be that decoherence between a post-quantum theory and quantum theory is very different from what our quantum/classical intuitions might lead us to believe.

The decoherence map of equation~\eqref{decmap} appears strange at first, since it maps states of a post-quantum system to a sub-system which is embedded in such a way that its states are not valid states of the post-quantum system. However the image of this decoherence map is actually the state space one would obtain if one had access to the post-quantum system $\ketbra{\psi}{\psi}^{\otimes 2}$  but only a restricted set of measurements. An observer with access to the post-quantum system $\ketbra{\psi}{\psi}^{\otimes 2}$ but only effects of the form $F(\psi) = {\rm Tr}((\hat F \otimes I)\ketbra{\psi}{\psi}^{\otimes 2})$ would reconstruct a quantum state space. Hence the link between the system and subsystem becomes clearer: the quantum sub-system is obtained from the post-quantum system by restricting the measurements on the post-quantum system. We emphasise once more 	that the toy model with unrestricted effects does not compose.

\section{Conclusion}

\subsection{Summary}

We have studied composition in general theories which have the same dynamical and compositional postulates as quantum theory but which have different measurement postulates. We presented a toy model of a bi-partite system with alternative measurement rules, showing that composition is possible in such theories. We showed that all such theories violate two compositional principles: local tomography and purification.

\subsection{Future work}

The toy model introduced in this work applies only to bi-partite systems and is simulable with quantum theory. Hence an important next step is constructing a toy model with alternative measurement rules which is consistent with composition of more than two systems. This requires a $\star$ product which is associative. This construction may prove impossible, or it may be the case that all valid constructions are simulable by quantum theory. We suggest that there are three possibilities when considering theories with fully associative products.

\bigskip

\bigskip\noindent{\it Possibility 1. (Logical consistency of postulates of Quantum Theory).}
  The only measurement postulates which are fully consistent with the associativity of composition are the quantum measurement postulates.
  
\bigskip
\noindent
If this were the case it would show that the postulates of quantum theory are not independent. Only the quantum measurement rules would be consistent with the dynamical and compositional postulates (and operationalism). However it may be the case that we can develop theories with alternative Born rules which compose with an associative product, but that all these theories are simulable by quantum theory.

\bigskip\noindent{\it Possibility 2. (Simulability of systems in theories with alternative postulates).}
  The only measurement postulates which are fully consistent with the dynamical postulates of quantum theory describe systems which are simulable with a finite number of quantum systems.
 
\bigskip\noindent{\it Possibility 3. (Non-simulability of systems in theories with alternative postulates).}
   There exist measurement postulates which are fully consistent with the dynamical postulates of quantum theory which describe systems which are not simulable with a finite number of quantum systems.
  
\bigskip
\noindent
This final possibility would be interesting from the perspective of GPTs as it would show that there are full theories which can be obtained by modifying the measurement postulates of quantum theory. It would show that quantum theory is not, in fact, an island in theory space.

\section{Acknowledgements}

We are grateful to Jonathan Barrett and Robin Lorenz for helpful discussions. We thank the referees for their comments which helped substantially improve the first version of this manuscript. TG is supported by the Engineering and Physical Sciences Research Council [grant number EP/L015242/1]. LM is funded by EPSRC.

\newpage

\bibliographystyle{unsrtnat}
\bibliography{refs}{}

\appendix

\onecolumn

\section{Operational principles and consistency constraints}\label{OPFCons}

\subsection{Single system}

As stated in the main section the allowed sets of OPFs $\mathcal F_d$ are subject to some operational constraints. We introduce features obeyed by operational theories and derive the consistency constraints {\bf C1. - C5.}  from them. In the following we adopt a description of operational principles in terms of circuits, as in~\cite{Chiribella_informational_2011}.
The basic operational primitives are preparations, transformations and measurements. These three are all procedures. We define them in terms of inputs, outputs and systems.

\begin{definition}[Preparation procedure]
	Any procedure which has no input and outputs one or more systems is a preparation procedure.
\end{definition}

\begin{figure}[H]
\begin{center}
\begin{tikzpicture}
\draw (0,0) rectangle (1,1) node[midway] {$\mathcal P$};
\draw (1,0.5) -- (2,0.5);
\end{tikzpicture}
\caption{Diagrammatic representation of a single system preparation procedure}
\end{center}
\end{figure}
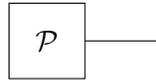

\begin{definition}[Transformation procedure]
	Any procedure which inputs one or more systems and outputs one or more systems is a transformation.
\end{definition}

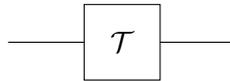
\begin{figure}[H]
\centering{
\begin{tikzpicture}
\draw (0,0) rectangle (1,1) node[midway] {$\mathcal T$};
\draw (1,0.5) -- (2,0.5);
\draw (-1,0.5) -- (0,0.5);
\end{tikzpicture}
}
\caption{Diagrammatic representation  of a single system transformation procedure.}
\end{figure}

\begin{definition}[Measurement procedure]
	Any procedure which inputs on or more systems and has no output is a measurement procedure.
\end{definition}

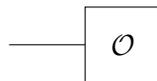
\begin{figure}[H]
\centering{
\begin{tikzpicture}
\draw (0,0) rectangle (1,1) node[midway] {$\mathcal O$};
\draw (-1,0.5) -- (0,0.5);
\end{tikzpicture}
}
\caption{Diagrammatic representation of a single system measurement procedure.}
\end{figure}

In the above ``no input" and ``no output" refers to output or input of systems, typically there will be a classical input or output such as a measurement read-out.

\begin{opimp}[Composition of a procedure with a transformation]
	The composition of a transformation with any procedure is itself a procedure of that kind. 
\end{opimp}
\noindent
For example the composition of a transformation and a measurement is itself a measurement.

\begin{opassum}[Mixing]
The process of taking two procedures of the same kind and implementing them probabilistically generates a procedure of that kind.
\end{opassum}

\begin{definition}[Experiment]
	An experiment is a sequence of procedures which has no input or output.
\end{definition}
\noindent
This principle entails that every experiment can be considered as a preparation and a measurement. The experiment is fully characterised by the probabilities
\begin{equation}
p(\mathcal O | \mathcal P) \ ,
\end{equation}
for all measurement outcomes $\mathcal O$ and all preparations $\mathcal P$ in the experiment.
\begin{figure}[!htb]
	\centering{
\begin{tikzpicture}
\draw (0,0) rectangle (1,1) node[midway] {$\mathcal P$};
\draw (1,0.5) -- (2,0.5);
\draw (2,0) rectangle (3,1) node[midway] {$\mathcal O$};
\end{tikzpicture}
}
\caption{All experiments can be represented as a preparation and a measurement.}
\end{figure}
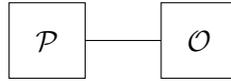

In this approach a system is an abstraction symbolised by the wire between the preparation procedure and the measurement procedure. A system $\sA$ can be represented as:
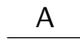
\begin{figure}[!htb]
	\centering{
\begin{tikzpicture}
\draw (1,0.5) -- node[above] {\sA} (2,0.5);
\end{tikzpicture}
}
\caption{Diagrammatic representation of a system.}
\end{figure}

\subsection{Pairs of systems}

Given the above definitions it is natural to ask when an experiment can be described using multiple systems. Let us consider a system  which we represent using two wires:
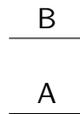
\begin{figure}[H]
\centering{
\begin{tikzpicture}
\draw (1,0.5) -- node[above] {\sA} (2,0.5);
\draw (1,1.5) -- node[above] {\sB} (2,1.5);
\end{tikzpicture}}
\caption{Diagrammatic representation of a pair of systems.}
\end{figure}
These can only be considered as representing two distinct systems $\sA$ and $\sB$ if it is possible to independently perform operations (transformations and measurements) on both systems. 

\begin{definition}[Existence of subsystems~\cite{Masanes_existence_2013}]
A system can be considered as a valid composite system if an operation on subsystem $\sA$ and an  operation on subsystem $\sB$ uniquely specify an  operation on $\sA \sB$ independent of the temporal ordering.
\end{definition}
\noindent
If the above property is not met, then the system cannot be considered as a composite (and should be represented using a single wire).
Diagrammatically this entails that any preparation of a composite system is such that:
\begin{figure}[H]
	\centering{
\begin{tikzpicture}
\draw (0,0) rectangle (1,2.5) node[midway] {$\mathcal P_{\sA \sB}$};
\draw (1,0.5) -- (2,0.5);
\draw (2,0) rectangle (3,1) node[midway] {$\mathcal O_\sB$};
\draw (4,1.5) rectangle (5,2.5) node[midway] {$\mathcal O_\sA$};
\draw (1,2) -- (4,2);
\draw (5.5,1.25) node{$=$};
\draw (6,0) rectangle (7,2.5) node[midway] {$\mathcal P_{\sA \sB}$};
\draw (7,0.5) -- (10,0.5);
\draw (10,0) rectangle (11,1) node[midway] {$\mathcal O_\sB$};
\draw (8,1.5) rectangle (9,2.5) node[midway] {$\mathcal O_\sA$};
\draw (7,2) -- (8,2);
\end{tikzpicture}
}
\caption{Definition of a composite system in diagrammatic form.}
\end{figure}
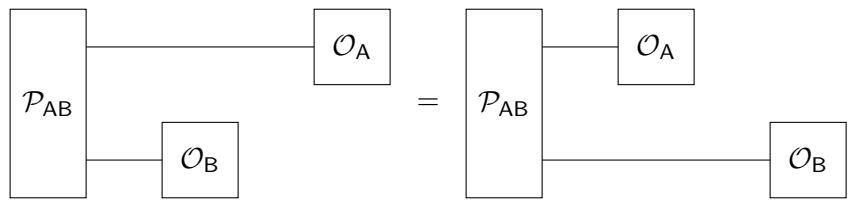
\begin{opimp}[Joint measurements]
	Every measurement outcome $\mathcal O_\sA$ on $\sA$ and $\mathcal O_\sB$ on $\sB$ defines a unique outcome $(\mathcal O_\sA , \mathcal O_\sB)$ on $\sA \sB$. 
\end{opimp}

The most general form of an experiment with two systems is:
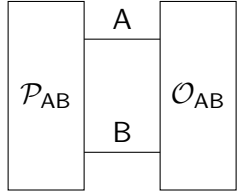
\begin{figure}[H]
\centering{
\begin{tikzpicture}
\draw (0,0) rectangle (1,2.5) node[midway] {$\mathcal P_{\sA \sB}$};
\draw (1,0.5) --  node[above] {\sB} (2,0.5);
\draw (2,0) rectangle (3,2.5) node[midway] {$\mathcal O_{\sA \sB}$};
\draw (1,2) --  node[above] {\sA} (2,2);
\end{tikzpicture}
}
\caption{Generic two system experiment.}
\end{figure}
\noindent
which can naturally be viewed as an experiment on a single system $\sA \sB$.

\begin{definition}[Separable procedures~\cite{Hardy_a_2010,Chiribella_probabilistic_2010}]
	Two independent preparations $\mathcal P_\sA$  and $\mathcal P_\sB$ which are independently measured with outcomes $\mathcal O_\sA$  and $\mathcal O_\sB$ are such that: 
\begin{equation*}
p(\mathcal O_\sA ,\mathcal O_\sB| \mathcal P_\sA , \mathcal P_\sB )   = p(\mathcal O_\sA | \mathcal P_\sA ) p(\mathcal O_\sB | \mathcal P_\sB ) \ , 
\end{equation*}
In this case the joint procedures $\mathcal P_\sA , \mathcal P_\sB$ and $\mathcal O	_\sA , \mathcal O_\sB$ are said to be separable.
\end{definition}

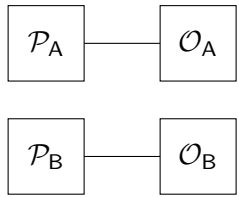
\begin{figure}[!htb]
	\centering{
\begin{tikzpicture}
\draw (0,0) rectangle (1,1) node[midway] {$\mathcal P_\sB$};
\draw (1,0.5) -- (2,0.5);
\draw (2,0) rectangle (3,1) node[midway] {$\mathcal O_\sB$};
\draw (0,1.5) rectangle (1,2.5) node[midway] {$\mathcal P_\sA$};
\draw (1,2) -- (2,2);
\draw (2,1.5) rectangle (3,2.5) node[midway] {$\mathcal O_\sA$};
\end{tikzpicture}
}
\caption{Separable two system experiment.}
\end{figure}

By the definition of a preparation, any operational procedure which outputs a system is a preparation. Hence consider the case where the measurement is separable. The procedure of making a joint preparation and making a measurement on $\sB$ is a preparation of a state $\sA$.

\begin{opimp}[Steering as preparation]
	Operationally Alice can make a preparation of system $\sA$ by making a preparation of $\sA \sB$ and getting Bob to make a measurement on system $\sB$.
\end{opimp}

\begin{figure}[!htb]
	\centering{
\begin{tikzpicture}
\draw (0,0) rectangle (1,2.5) node[midway] {$\mathcal P_{\sA \sB}$};
\draw (1,0.5) -- (2,0.5);
\draw (2,0) rectangle (3,1) node[midway] {$\mathcal O_\sB$};
\draw (1,2) -- (4,2);
\end{tikzpicture}
}
\caption{Preparation by steering.}
\end{figure}
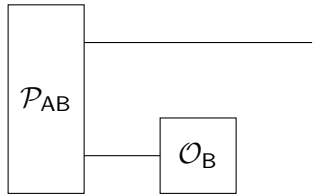

By the definition of a measurement any operational procedure which inputs a system and outputs no system is a measurement. Consider the case where the preparation is separable. Then the procedure of preparing system $\sB$ and jointly measuring $\sA$ and $\sB$ is a measurement procedure on $\sA$.

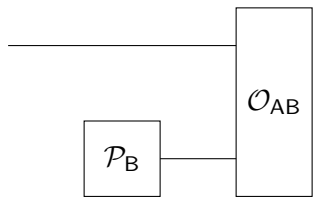
\begin{figure}[!htb]
	\centering{
\begin{tikzpicture}
\draw (0,0) rectangle (1,1) node[midway] {$\mathcal P_\sB$};
\draw (1,0.5) -- (2,0.5);
\draw (2,0) rectangle (3,2.5) node[midway] {$\mathcal O_{\sA \sB}$};
\draw (-1,2) -- (2,2);
\end{tikzpicture}
}
\caption{Measurement using ancilla.}
\end{figure}

\begin{opimp}[Measuring with an ancilla]
	A valid measurement for Alice consists in adjoining her system to an ancillary system $\sB$ and carrying out a joint measurement.
\end{opimp}
\noindent
In the case where both preparation and measurement are separable then the experiment can be viewed as two separate experiments. 
In the bi-partite case there are no further methods of generating preparations and measurements. Hence when determining whether a pair of systems is consistent with the operational properties which arise from composition, these are the only features which we need to consider.

One may ask whether any further operational implications will emerge from considering more than two systems.
\begin{opassum}[Associativity of composition]
	The systems $(\sA \sB) \sC$ and $\sA (\sB \sC)$ are the same. 
\end{opassum}
\noindent
This implies that there are no new types of procedures which can be carried out on a single system by appending more than one system. Consider an experiment with multiple systems $\sA, \sB , \sC... $. For any partitioning of the experiment which creates a preparation of system $\sA$, all regroupings of systems $\sB, \sC , ...$ are equivalent. This is a preparation by steering of system $\sA$ conditional on a measurement on systems $\sB , \sC , ...  $ which can be viewed as a single system. Diagrammatically it tells us that all ways of partitioning an experiment with multiple systems are equivalent.

This entails the only constraints imposed by the operational framework will come from the assumptions and implications outlined above. There are no further operational implications which emerge from the above definitions and assumptions.

In the next section we translate the operational features above into the language of OPFs, and show which constraints they impose on the OPF sets $\F_d$.

\subsection{Consistency constraints}

We assume the Finiteness Principle holds, and that for a set of OPFs $\mathcal F_d$ there exists a finite linearly generating set $\{F_i\}_{i = 1}^{K_d}$. That is to say:
\begin{equation}
F= \sum_i c_i F_i \ .
\end{equation}

\subsubsection{Constraint C1}

Consistency constraint {\bf C1} follows directly from the fact that the composition of a transformation and a measurement is a measurement.

\subsubsection{Constraint C2 }

\begin{definition}[State]
A state corresponds to an equivalence class of indistinguishable preparation procedures. 
\end{definition}
\noindent
From this definition it follows that two states cannot be indistinguishable. This implies {\bf C2}. In a system where some pure states are indistinguishable the manifold of pure states would no longer be the set of rays on $\mathbb C^d$ (as required by the first postulate).

\subsubsection{Constraint C3 }

Consider a composite system $\sA \sB$. By the definition of a composite system above, for any OPF $F_\sA$ on $\sA$ and $F_\sB$ on $\sB$ there exists an OPF $F_\sA \star F_\sB$ on $\sA \sB$. By the assumption that mixing is possible, the outcome $\{p_i, F_\sA^i\}$ is a valid outcome. 
\begin{opassum}[Mixing separable procedures]
	If two parallel processes are separable, it is equivalent to mix them before they are considered as a joint process or after. 
\end{opassum}
From the above assumption it follows that:
\begin{align}
(\sum_i p_i F_\sA^i) \star F_\sB  = \sum_i p_i (F_\sA^i \star F_\sB) \\
F_\sA \star(\sum_i p_i   F_\sB^i) = \sum_i p_i (F_\sA \star F_\sB^i) 
\end{align}
If we further assume that it is possible to mix with subnormalised probabilities, i.e. $\sum_i p_i \leq 1$ then the $\star$ product is bi-linear.
The identity ${\bf u}_\sA \star {\bf u}_\sB = {\bf u}_{\sA \sB}$ follows from the fact that a separable measurement is a valid measurement on $\sA \sB$, and hence:
\begin{equation}
{\bf u}_{\sA \sB} = \sum_{i,j} (F_\sA^i \star F_\sB^j) = {\bf u}_\sA \star {\bf u}_\sB  \ .
\end{equation}

\begin{principle}[Uncorrelated pure states]
Given two systems $\sA$ and $\sB$ independently prepared in pure states $ \psi_\sA$ and  $ \phi_\sB$ the joint state of the system $\sA \sB$ is given by $\psi_\sA \otimes \phi_\sB$. 
\end{principle}

This principle, together with the definition of independent systems implies that:
\begin{equation}
(F_\sA \star F_\sB)(\psi_\sA \otimes \phi_\sB) = F_\sA(\psi_\sA) F_\sB(\phi_\sB)
\end{equation}

\subsubsection{Constraint C4 }

Let us consider the steering scenario. In this case the process of both Alice and Bob making a measurement with outcome $F_\sA \star F_\sB$ on a joint state $\phi_{\sA \sB}$ can be considered as a measurement with outcome $F_\sA$ on system $\sA$ prepared in a certain manner. 

An arbitrary preparation of system $\sA$ is given by $\{p_i , \psi^i_\sA \}$. Hence the steering preparation implies that for each preparation of $\sA \sB$  and for each local measurement outcome on $\sB$ there exists a state $\{p_i , \psi^i_\sA \}$ in which system $\sA$ is prepared.  In the OPF formalism this means that for every $\phi_{\sA\sB} \in \mathbb{C}^{d_\sA d_\sB}$ and every $F_\sB \in \mathcal F_{d_\sB}$ there exists an ensemble  $\{p_i \phi^i_\sA\}$ such that
\begin{equation}
\frac{(F_\sA \star F_\sB)(\phi_{\sA \sB})}{({\bf u}_\sA \star F_\sB)(\phi_{\sA \sB})} = \sum_i p_i F_\sA (\psi^i_\sA) \ ,
\end{equation}
holds for all $F_\sA \in \mathcal F_{d_\sA}$. The normalisation occurs due to the fact that summing over the measurement outcomes $F_\sA$ should give unity on both sides of the expression. 

\subsubsection{Constraint C5}

Let us consider the scenario consisting in measuring with an ancilla. In this case Alice and Bob carry out a joint measurement with outcomes $F_{\sA \sB}$ on a system in an uncorrelated state $\psi_\sA \otimes \phi_\sB$. This should correspond to a valid measurement with outcome $F_\sA'$ on system $\sA$ prepared in state $\psi_\sA$. For each $F_{\sA \sB} \in \mathcal F_{d_\sA d_\sB}$ and for each $\phi_\sB \in \mathbb C^{d_\sB}$ there exists an $F_\sA' \in \mathcal F_{d_\sA}$ such that
\begin{equation}
F_{\sA \sB}(\psi_\sA \otimes \phi_\sB) = F_\sA'(\psi_\sA) \ ,
\end{equation}
for all $\psi_\sA \in \mathbb C^{d_\sA}$.

\section{Violation of purification}

In this appendix we show that all alternative measurement postulates violate purification (for an arbitrary choice of ancillary system dimension).

As shown in \cite{Galley_classification_2017} the representations $\Gamma^{d}$ corresponding to alternative measurement postulates for systems with pure states $\pcd$ ($d>2$) are of the form 
\begin{equation}\label{generalrep}
\Gamma =  \bigoplus_{j\in \mathcal J} 
\mathcal D^d_j\ ,
\end{equation}
where $\mathcal J$ is a list of non-negative integers (at least one of which is not 0 or 1) and $\mathcal D_j^d$ are representations of $\sud$ labelled by Young diagrams $(2j,  \underbrace{j,  \ldots,j}_{d-2})$.

Consider a system $\mathcal S_{\sA\sB} = \{\pcdAdB , \Gamma_{\sA\sB} \}$ which is the composite of two systems $S_{\sA} = \{\pcdA , \Gamma_{\sA} \}$ and $S_{\sB} = \{\pcdB , \Gamma_{\sB} \}$. Here the representations $\Gamma$ are of the form \eqref{generalrep}. Let us define the following equivalence classes of pure global states:
\begin{equation}
[\ket{\psi}_{\sA\sB}]_{U_\sB} =  \{ \ket{\psi}_{\sA\sB}' \in \pcdAdB |\ket{\psi}_{\sA\sB}' =  \unity_\sA \otimes U_\sB \ket{\psi}_{\sA\sB}  \}
\end{equation}
All members of the same equivalence class are necessarily mapped to the same reduced state of Alice. Otherwise, Bob could signal to Alice. We note that~\cite{Barnum_no_2003} makes use of this observation in a similar context.  Let us call the set of all these equivalence classes $R_\sB$.
\begin{equation}
R_\sB := \{ [\ket{\psi}_{\sA\sB}]_{U_\sB} | \ket{\psi}_{\sA\sB} \in \mathbb C^{d_\sA d_\sB} \}
\end{equation}
The map from global states to reduced states can be defined on the equivalence classes $ [\ket{\psi}_{\sA\sB}]_{U_\sB}$ since two members of the same equivalence class are always mapped to the same reduced states.  $\mathcal R: R_\sB \rightarrow \mathcal S_\sA$ is the map from equivalence classes to reduced states:
\begin{equation}
\mathcal R([\ket{\psi}_{\sA\sB}]_{U_\sB}) = \bar \omega_\sA (\ket{\psi}_{\sA\sB}) \ ,
\end{equation}
where $\bar \omega_\sA (\ket{\psi}_{\sA\sB})$ is the reduced state obtained in the standard manner from the global state $\ket{\psi}_{\sA\sB}$ (as outlined in the following appendix).
Next we prove that the image of $\mathcal R$ is smaller than $\mathcal S_\sA$ for any non-quantum measurement postulates. 
In other words there are some (local) mixed states in $\mathcal S_\sA$ which are not reduced states of the global pure states $\ket \psi_{\sA \sB}$.

In the Schmidt decomposition a state $\ket{\psi}_{\sA\sB}$ is:
\begin{equation}
\ket{\psi}_{\sA\sB} = \sum_{i=1}^{d_A} \lambda_i \ket{i}_\sA \ket{i}_\sB  , \lambda_i \in \mathbb R  \ , \ \sum_i \lambda_i^2 = 1 \ ,
\end{equation}
where we assume that the Schmidt coefficients are in decreasing order $\lambda_i \geq \lambda_{i+1}$.
Two states with the same coefficients and the same basis states on Alice's side belong to the same equivalence class $[\ket{\psi}_{\sA\sB}]_{U_\sB}$.  
Also, two Alice's basis differing only by phases (e.g.~$\{ \ket i_\sA \}$ and $\{ e^{i \theta_i}\! \ket i_\sA \}$) give rise to the same equivalence class. Because the phases $e^{i \theta_i}$ can be absorbed by Bob's unitary.

Let us count the number of parameters that are required to specify an equivalence class in $R_\sB$. First, we have the $d_A-1$ Schmidt coefficients. Second, we note that the number of parameters to specify a basis in $\mathbb C^{d_\sA}$ is the same as to specify an element of U$(d_\sA)$. Which is the dimension of its Lie algebra,  $d_\sA^2$, the set of anti-hermitian matrices.
Third, we have to subtract the $d_\sA$ irrelevant phases $\theta_i$. The three terms together give
\begin{equation}
(d_\sA -1)+ d_\sA^2 -d_\sA
= d_\sA^2-1
\end{equation} 
Hence $d_\sA^2 -1$ parameters are needed to specify elements of $R_\sB$. The set ${\rm Image}(\mathcal R)$ requires the same or fewer parameters to describe as $R_\sB$. This follows from the fact that every element of $R_\sB$ can be mapped to distinct images, or multiple elements can be mapped to the same image.

Hence by requiring that Alice's reduced states are in one-to-one correspondence with these equivalence classes, her state space must have a dimension $d_\sA^2 - 1$. The only measurement postulates which generate a state space with this dimension are the quantum ones. This follows from the fact that the dimension of the irreducible representations $\D_j^d$ corresponding to alternative measurement postulates are given by:
\begin{equation}\label{dimension}
D_j^d = \left(\frac{2j}{d-1}+1\right) \prod_{k=1}^{d-2} 
\left(1 +\frac j k \right)^2
\ ,
\end{equation}
This is equal to $d^2 - 1$ for the case $j=1$ (corresponding to the quantum state space). Moreover, since this is the lowest dimensional (non-trivial) such representation there are no reducible representations of the form $\bigoplus_i \D_i^d$ which are of dimension $d^2 -1$.

\section{OPF formalism, representation theory and local tomography}\label{OPFformalism}

In this appendix we show that each set of OPFs $\mathcal F_d$ is associated to a representation of the group $\sud$. We also show that the representations associated to locally tomographic systems with sets of OPFs $\mathcal F_{d_\sA d_\sB}$ have certain features when restricted to the local subgroup $\sudA \times \sudB$. It is these features which will be used in the next appendix to show that all non-quantum measurement postulates lead to a violation of local tomography.
In the following we assume the Finiteness Principle holds.

\subsection{Single systems}

\begin{lemma}
	To each set of measurement postulates $\mathcal F_d$ obeying the Finiteness Principle there exists a unique representation $\Gamma^d$ of $\sud$ associated to that set. 
\end{lemma}

\begin{proof}
	
	We take a set of measurement postulates $\mathcal F$, where $\{F_i\}$ form a basis for $\mathcal F$.
	\begin{equation}
	F(\psi) = \sum_i c_i F_i (\psi) \ , \forall \psi \ .
	\end{equation}
	We consider an OPF $F \circ U$:
	\begin{equation}
	(F \circ U)(\psi) = F(U \psi) =  \sum_i c_i F_i(U \psi)
	\end{equation}
	Where
	\begin{equation}
	F_i(U \psi) = (F_i \circ U) (\psi) = \sum_j \bar \Gamma_i^j(U) F_j (\psi) \ .
	\end{equation}
	Hence
	\begin{equation}
	F \circ U (\psi) = \sum_{ij} c_i \bar \Gamma_i^j(U) F_j(\psi) 
	\end{equation}
	Consider
	\begin{align}
	F   ( U U' \psi) &  =  \sum_{ij} c_i \bar \Gamma_i^j(U) F_j(U'\psi) \nonumber  \\  & =   \sum_{ijk} c_i \bar \Gamma_i^j(U) \bar \Gamma_j^k(U') F_k(\psi) 
	\end{align}
	We can also consider $UU'$ as a single element:
	\begin{equation}
	F   ( U U' \psi) = \sum_{ik} c_i \bar \Gamma_i^k(UU') F_k(\psi) 
	\end{equation}
	This shows that $\bar \Gamma(UU') = \bar \Gamma(U)\bar \Gamma(U')$ and the map $\bar \Gamma: U \mapsto \bar \Gamma(U)$ is a representation of $\sud$.
\end{proof}

\begin{lemma}
	The representation $\bar \Gamma^d$ of measurement postulates $\mathcal F_d$ contains a unique trivial subrepresentation.
\end{lemma}
\begin{proof}
	Consider a set of OPFs $\{F^{(i)}\}$ which form a measurement. The OPF $\mathbf u = \sum F^{(i)}$ is such that $\mathbf u (\psi) = 1 \ \forall \psi$. Consider the basis $\{F_i\}$ where $F_1 = \mathbf u$. We observe that $\mathbf u (\psi) = \mathbf u ( U \psi) , \ \forall U \ \forall \psi$. This implies that $\bar \Gamma(U) \mathbf u = \mathbf u \  \forall U$ and that the representation $\Gamma$ has a trivial component. If the representation had another trivial component,  it would necessarily be linearly dependent on the first. It would then be a redundant entry in the list of fiducial outcomes which is contrary to the property that they are linearly independent.
\end{proof}

\subsection{Composite systems}

The measurement structure $\mathcal F_{\sA \sB}$ contains all measurements of the form $F_\sA \star F_\sB$ where $F_\sA \in \mathcal F_\sA $ and $F_\sB \in \mathcal F_\sB$. Let $\{F_\sA^i\}$ and $\{F_\sB^j\}$ be bases for the two OPF spaces. Then the OPFs $F_\sA^i \star F_\sB^j$ form a basis for the global OPFs $F_\sA \star F_\sB$. Hence a basis for $\mathcal F_{\sA\sB}$ is $\{F_\sA^i \star F_\sB^j , F_{\sA\sB}^k \}$~\cite{Hardy_foliable_2009, Masanes_lecture_2017}. 

\subsubsection{Local tomography}

A bi-partite system is locally tomographic if $\{F_\sA^i \star F_\sB^j\}_{ij}$ is a basis for $\R \F_{d_\sA d_\sB}$.

\begin{lemma}
	For a locally tomographic bi-partite system with representation $\bar \Gamma^{d_\sA d_\sB}$ the restriction of $\bar \Gamma^{d_\sA d_\sB}$ to $\sudA \times \sudB$ is:
	\begin{equation}
	\bar \Gamma^{d_\sA d_\sB}_{|\sudA \times \sudB} = \bar \Gamma^{d_\sA} \boxtimes \bar \Gamma^{d_\sB}
	\end{equation}
\end{lemma}

\begin{proof}
	Let us consider the action of an element of $\sudA \times \sudB$ on an OPF $F_{\sA \sB} = \sum_{ij} \gamma_{ij} (F_\sA^i \otimes F_\sB^j)$ in $\mathcal F_{\sA \sB}$ .
	\begin{align}
	\bar \Gamma^{d_\sA d_\sB}_{U_{\sA} \otimes U_{\sB}} F_{\sA \sB} = F_{\sA \sB} \circ (U_\sA \otimes U_\sB) 
	=  \sum_{ij} \gamma_{ij} (F_\sA^i \otimes F_\sB^j) \circ (U_\sA \otimes U_\sB)  \ .
	\end{align}
	Using the fact that $(F_{\sA} \otimes F_{\sB}) \circ (U_\sA \otimes U_\sB) = (F_\sA  \circ  U_\sA) \otimes (F_\sB \circ U_\sB)$:
	\begin{equation}
	\bar \Gamma^{d_\sA d_\sB}_{U_{\sA} \otimes U_{\sB}} F_{\sA \sB} =  \sum_{ij} \gamma_{ij} (F_\sA^i  \circ  U_\sA) \otimes (F_\sB^j \circ U_\sB) \ .
	\end{equation}
	From the actions of $\sudA$ and $\sudB$ on $\R \F_{d_\sA}$ and $\R \F_{d_\sB}$ we have:
	\begin{align}
	& F_\sA^i  \circ  U_\sA  = \bar \Gamma^{d_\sA}_{U_A}  F_\sA^i \ , \\ 
	& F_\sB^j  \circ  U_\sB  =  \bar \Gamma^{d_\sB}_{U_B} F_\sB^j \ . 
	\end{align}
	Hence,
	\begin{align}
	\bar \Gamma^{d_\sA d_\sB}_{U_{\sA} \otimes U_{\sB}} F_{\sA \sB}   = \sum_{ij} \gamma_{ij} (\bar \Gamma^{d_\sA}_{U_A}  F_\sA^i  \otimes   \bar \Gamma^{d_\sB}_{U_B} F_\sB^j)  = \sum_{ij} \gamma_{ij} (\bar \Gamma^{d_\sA}_{U_A} \otimes   \bar \Gamma^{d_\sB}_{U_B}) ( F_\sA^i  \otimes F_\sB^l)  =  (\bar \Gamma^{d_\sA}_{U_A} \otimes \bar \Gamma^{d_\sB}_{U_B}) F_{\sA \sB} \ .
	\end{align}
\end{proof}

\subsubsection{Holistic systems}

A bi-partite system which is not locally tomographic is \emph{holistic}. Real vector space quantum theory is an example of a holistic theory~\cite{Hardy_limited_2012}.
A basis for $\mathcal F_{d_\sA d_\sB}$ in a holistic bi-partite system is $\{F_\sA^i \star F_\sB^j , F_{\sA\sB}^k \}_{ijk}$~\cite{Hardy_foliable_2009,Masanes_lecture_2017}. Here $\{F_\sA^i \star F_\sB^j\}_{ij}$ span the locally tomographic subspace of $\F_{d_\sA d_\sB}$ denoted $\F_{d_\sA d_\sB}^{\rm LT}$. Due to bilinearity of the $\star$ product the map $\star: \R \F_{d_\sA} \times \R \F_{d_\sB} \to \R \F_{d_\sA d_\sB}^{\rm LT}$ is isomorphic to a tensor product.

\begin{lemma}
	For a holistic bi-partite system with representation $\bar \Gamma^{d_\sA d_\sB}$ the restriction of $\bar \Gamma^{d_\sA d_\sB}$ to $\sudA \times \sudB$ is:
	\begin{equation}
	\bar \Gamma_{|\sudA \times \sudB}^{d_\sA d_\sB} =  \bar \Gamma^{d_\sA} \boxtimes \bar \Gamma^{d_\sB} \oplus \bigoplus_i  \Gamma_i^{d_\sA} \boxtimes  \Gamma_i^{d_\sB} \ ,
	\end{equation}
	where the representations $\bar \Gamma^{d_\sA d_\sB}, \bar \Gamma^{d_\sA}$ and $\bar \Gamma^{d_\sB}$ contain a trivial representation. This is not necessarily the case for $\Gamma_i^{d_\sA}$ and $\Gamma_i^{d_\sB}$ (which may not be of the form $\D_j^{d_\sA}$ or $\D_j^{d_\sB}$).
\end{lemma}

\begin{proof}
In holistic systems a basis for $\mathcal F_{\sA\sB}$ is $\{F_\sA^i \otimes F_\sB^j , F_{\sA\sB}^k \}_{ijk}$. 
\begin{equation}
F_{\sA \sB} = \sum_{ij} \gamma_{ij}^{{\rm LT}} (F_\sA^i \otimes F_\sB^j) + \sum_k  \gamma_{k}^{{\rm H}} F^k_{\sA \sB}  = F^{\rm LT}_{\sA \sB} + F^{\rm H}_{\sA \sB} \ .
\end{equation}
We consider the action of a $\sudA \times \sudB$ subgroup on ${\rm span}(\{F_\sA^i \otimes F_\sB^j\}_{ij})$.
\begin{equation}
F_\sA^i \otimes F_\sB^j \circ (U_\sA \otimes U_\sB) = (F_\sA^i \circ U_\sA) \otimes (F_\sB^j \circ U_\sB) \ .
\end{equation}
The action of $\sudA \times \sudB$ maps basis elements of the form $F_\sA \otimes F_\sB$ to other elements of that form. Hence ${\rm span}(F_\sA^i \otimes F_\sB^j)$ is a proper subspace of $\mathcal F_{\sA\sB}$ left invariant under the action of $\sudA \times \sudB$. The representation $ \bar \Gamma_{|\sudA \times \sudB}^{d_\sA d_\sB} $ is reducible and decomposes as:
\begin{equation}
\bar \Gamma_{ |\sudA \times \sudB}^{d_\sA d_\sB}  = \bar \Gamma_{{\rm LT} |\sudA \times \sudB}^{d_\sA d_\sB}  \oplus \Gamma_{{\rm H}|\sudA \times \sudB}^{d_\sA d_\sB} \ .
\end{equation}
The action $\bar \Gamma_{{\rm LT} |\sudA \times \sudB}^{d_\sA d_\sB}$ on the locally tomographic subspace is of the form  $\bar \Gamma^{d_\sA} \boxtimes \bar \Gamma^{d_\sB}$ (as determined in the previous lemma). 	 $ \Gamma_{{\rm H}|\sudA \times \sudB}^{d_\sA d_\sB}$ is an arbitrary representation of $\sudA \times \sudB$ hence of the form $\bigoplus_i  \Gamma_i^{d_\sA} \boxtimes  \Gamma_i^{d_\sB}$.
\end{proof}

\subsection{Representation theoretic criterion for local tomography}

Consider a bi-partite system with alternative measurement postulates $\mathcal F_{d_\sA d_\sB}$, $\mathcal F_{d_\sA}$ and $\mathcal F_{d_\sB}$. If the theory is locally tomographic then necessarily: 
\begin{equation}\label{linlocaltom}
\bar \Gamma_{| \sudA \times \sudB}^{d_\sA d_\sB} = \bar \Gamma_{\sA} \boxtimes \bar \Gamma_\sB
\end{equation}
Where $\bar \Gamma_{\sA}$ and $\bar \Gamma_\sB$ both contain a unique trivial representation. By contraposition, any system  which has a representation $\bar \Gamma_{\sA\sB}^{d_\sA d_\sB}$ which does not have this form when restricted $ \sudA \times \sudB$ cannot obey local tomography. This allows us to establish the following test for violation of local tomography:

\begin{lemma}
	Given a bi-partite system with measurement postulates $\mathcal F_{d_\sA d_\sB}$, $\mathcal F_{d_\sA}$ and $\mathcal F_{d_\sB}$. The associated representation are $\bar \Gamma_{\sA\sB}^{d_\sA d_\sB} $, $\bar \Gamma_{\sA}^{d_\sA}$ and $ \bar \Gamma_{\sB}^{d_\sB}$. If $\bar \Gamma_{| \sudA \times \sudB}^{d_\sA d_\sB}$ is not of the form \eqref{linlocaltom} then the system is holistic.
\end{lemma}

\subsubsection{Affine representation and existence of trivial times trivial as a criterion}

The above lemmas can be translated into the affine representation  $\Gamma^d$  which do not contain a trivial subrepresentation. 
\begin{equation}
\bar  \Gamma^{d}  = 1^{d} \oplus  \Gamma^{d} \ .
\end{equation}
where $1^d$ is the trivial representation of $\sud$. We can decompose the tensor product action:
\begin{align}
\bar  \Gamma^{d_\sA}  \boxtimes  \bar \Gamma^{d_\sB}  =  (1^{d_\sA} \oplus  \Gamma^{d_\sA}) \boxtimes (1^{d_\sB} \oplus  \Gamma^{d_\sB})
= (1^{d_\sA} \boxtimes 1^{d_\sB}) \oplus (1^{d_\sA} \boxtimes   \Gamma^{d_\sB}) \oplus ( \Gamma^{d_\sA} \boxtimes 1^{d_\sB}) \oplus ( \Gamma^{d_\sA} \boxtimes  \Gamma^{d_\sB}) \ .
\end{align}
The first term occurs due to the trivial component in $\bar \Gamma_{\sA\sB}^{d_\sA d_\sB} = 1^{d_\sA d_\sB} \oplus  \Gamma_{\sA\sB}^{d_\sA d_\sB}$.

For a locally tomographic system with OPF set $\mathcal F_{d_\sA d_\sB}$ and representation $\bar \Gamma^{d_\sA d_\sB}$, the restriction of $  \Gamma^{d_\sA d_\sB}$ to the local subgroup $\sudA \times \sudB$, has the following form:
\begin{equation}\label{localtomography}
\Gamma_{| \sudA \times \sudB}^{d_\sA d_\sB} =    (1^{d_\sA} \boxtimes   \Gamma^{d_\sB}) \oplus ( \Gamma^{d_\sA} \boxtimes 1^{d_\sB}) \oplus ( \Gamma^{d_\sA} \boxtimes  \Gamma^{d_\sB}) \ .
\end{equation}
This shows that for a locally tomographic theory the representations $ \Gamma_{\sA\sB| \sudA \times \sudB}^{d_\sA d_\sB} $ cannot contain any terms $1^{d_\sA} \boxtimes 1^{d_\sB}$. By contraposition we establish:
\begin{lemma}\label{lemnoloctom}
	Given a bi-partite system  $(\C^{d_{\sA\sB}} , \F_{d_\sA d_\sB}, \Gamma^{d_\sA d_\sB})$ with subsystems $(\C^{d_\sA} , \F_{d_\sA},\Gamma^{d_\sA})$ and $(\C^{d_\sB} , \F_{d_\sB}, \Gamma^{d_\sB})$ then if $ \Gamma^{d_\sA d_\sB}_{\sA \sB}$ has a subrepresentation $1^{d_\sA} \boxtimes 1^{d_\sB}$ upon restriction to $\sudA \times \sudB$ the composite system is holistic.
\end{lemma}

\section{Background representation theory}

\subsection{Background}

\subsubsection{Young diagram}

A Young diagram is a collection of boxes in left-justified rows, where each row-length is in non-increasing order. The number of boxes in each row determine a partition of the total number of boxes $n$. This partition denoted $\lambda$ is associated to the Young diagram which is said to be of shape $\lambda$. We write $|\lambda| = n$ for the total number of boxes.

A Young diagram with $m$ rows (labelled $1$ to $m$) where row $i$ has $n_i$ boxes (by definition $n_1 \geq n_2 \geq ... \geq n_m$) is written $\lambda = (n_1 , n_2 , ... , n_m)$. By definition:
\begin{equation}
\sum_i n_i = n
\end{equation}
\begin{equation}
\lambda^1 = \yng(2,1) \ , \quad \lambda^2 = \yng(3,3,2,1)  \ , \quad \lambda^3 =	\yng(5,2,1,1) \ .
\end{equation}
The above Young diagrams are $\lambda^1 = (2,1)$, $\lambda^2 = (3,3,2,1)$ and $\lambda^3 = (5,2,1,1)$. 

In the case where a diagram has $m$ multiple rows of the same length $l$ we write $l^m$ instead of writing out ``$l$" $m$ times.. For instance $\lambda^2 = (3^2,2,1)$

\subsubsection{Representations of $\sud$}

The special unitary group $\sud$ is the Lie group of $d \times d$ unitary matrices of determinant $1$. A finite-dimensional representation of $\sud$ is a smooth homomorphism $\sud \rightarrow \mathrm{GL}(V)$, with $V$ a finite-dimensional (complex) vector space. 

Irreducible representations of $\sud$ are labelled by Young diagrams with at most $d$ rows. There is no limit on the total number of boxes. A column of $d$ boxes can be removed from a Young diagram of a representation of $\sud$ without changing the representation labelled by the diagram. Hence an irreducible representation of $\sud$ corresponds to an equivalence class of Young diagrams up to columns of size $d$. Typically the canonical representative member of the equivalence class is the Young diagram where all columns of length $d$ have been removed.  With this choice of representative element of each class the Young diagram of representations of $\sud$ have at most $d-1$ rows. For example the equivalence class of rectangular Young diagrams of columns of length $d$ are all mapped to the empty diagram (since all columns are removed). 

\begin{equation}
\lambda^1 = \yng(4,3,3,2) \ , \quad \lambda^2 =   \yng(3,2,2,1) \ , \quad \lambda^3 = \yng(2,1,1) \ .
\end{equation}

These three Young diagrams are equivalent up to columns of length 4. They all label the same representation of $\mathrm{SU}(4)$. The canonical representative of the equivalence class is $\lambda^3$. In the following the expression ``representation of $\sud$ labelled by the Young diagram $\lambda$" is used to mean ``representation of $\sud$ labelled by the equivalence class of Young diagrams with representative member $\lambda$". The young diagram {\tiny \yng(1)} labels the fundamental representation of $\sud$. The empty diagram labels the trivial representation.

Given a canonical representative $\lambda$ we call $\lambda_f$ the diagram in the same equivalence class which has $f$ boxes (in other words to which a number of columns of length $d$ have been added such that the total number of boxes is $f$). For example $\lambda^2 = \lambda^3_8$.

$\Gamma_\lambda^d$ is the representation of $\sud$ associated to the Young diagram $\lambda$. The index $d$ will be dropped when the context makes it obvious.

\subsubsection{Representations of the symmetric group}

The symmetric group on $n$ symbols $S_n$ has as group elements all permutation operations on $n$ distinct symbols. The conjugacy classes of $S_n$ are labelled by partitions of $n$. Hence the number of (inequivalent) irreducible representations of $S_n$ is the number of partitions of $n$. Moreover these irreducible representations can be parametrised by partitions of $n$. These are labelled by Young diagrams with $n$ boxes \cite{Robinson_representation_1991}.
 
$\Delta_\lambda^n$ denotes the representation of $S_n$ labelled by the Young diagram $\lambda$. The index $n$ will be dropped when the context makes it obvious.

\subsection{Branching rule $\mathrm{SU}(mn) \rightarrow \mathrm{SU}(m) \times \mathrm{SU}(n)$}

\subsubsection{Definition}

Consider an irreducible representation $\Gamma_\lambda^{mn}$ of $\mathrm{SU}(mn)$ and restrict it to a $\mathrm{SU}(m) \times \mathrm{SU}(n)$ subgroup:
\begin{equation}
\Gamma^{mn}_\lambda(U_1 \otimes U_2)  , \forall U_1 \in \mathrm{SU}(m) \ , \forall U_2 \in \mathrm{SU}(n) \ .
\end{equation}
In general this will yield a reducible representation of $\mathrm{SU}(m) \times \mathrm{SU}(n)$. This representation will be built of irreducible representations  $\Gamma_\mu^m \boxtimes  \Gamma_\nu^{n}$
\begin{equation}
  (\Gamma_\mu^m \boxtimes  \Gamma_\nu^n) (U_1,U_2) = \Gamma_\mu^m(U_1) \otimes \Gamma_\nu^n(U_2) \ ,
\end{equation}
We write $\Gamma^{mn}_{\lambda| \mathrm{SU}(m) \times \mathrm{SU}(n)}$ for the restriction of $\Gamma_\lambda^{mn}$ to a $\mathrm{SU}(m) \times \mathrm{SU}(n)$ subgroup. In general:
\begin{equation}
\Gamma^{mn}_{ \lambda | \mathrm{SU}(m) \times \mathrm{SU}(n)} = \bigoplus_{\mu , \nu} \Gamma^m_\mu \boxtimes  \Gamma^n_\nu \ ,
\end{equation}
Where there can be repeated copies for a given $\mu , \nu$. In general finding which representations $\Gamma^m_\mu \boxtimes  \Gamma^n_\nu$ occur in this restriction is a hard problem. In the following we outline a method which allows us to determine the multiplicity of $\Gamma^m_\mu \boxtimes  \Gamma^n_\nu$ in $\Gamma^{mn}_{\lambda| \mathrm{SU}(m) \times \mathrm{SU}(n)}$. $\lambda$ , $\mu$ and $\mu$ will refer to the Young diagrams of $\Gamma^{mn}_\lambda$, $\Gamma^m_\mu$ and $\Gamma^n_\nu$ respectively. 

\subsubsection{Inner product of representations of the symmetric group}

We consider two representations $\Delta_\mu$ and $\Delta_\nu$ of $S_f$. We construct the Kronecker product of the two matrices $\Delta_\mu(s)$ and $\Delta_\nu(s)$ for all $s \in S_f$. This creates a representation which we call the tensor product  $\Delta_\mu \otimes \Delta_\nu$ (sometimes called the inner product). In general this is a reducible representation: 
\begin{equation}
\Delta_\mu \otimes \Delta_\nu = \bigoplus_\lambda g(\mu, \nu , \lambda) \Delta_\lambda
\end{equation}
Here we abuse notation slightly to mean that $g(\mu, \nu , \lambda)$ is the multiplicity of $\Delta_\lambda$ in $\Delta_\mu \otimes \Delta_\nu $. These $g(\mu, \nu , \lambda)$ are known as the Clebsch-Gordan coefficients of the symmetric group, and understanding them remains one of the main open problems in classical representation theory. These coefficients are also relevant in quantum information theory, as they are related to the spectra of statistical operators \cite{Christandl_Nonzero_2007}.

\subsubsection{Recipe}

What is the multiplicity of  $\Gamma^m_\mu \boxtimes  \Gamma^n_\nu$ in the restriction of  $\Gamma^{mn}_\lambda$ to $ \mathrm{SU}(m) \times \mathrm{SU}(n)$? We adopt the approach from~\cite{itzykson_unitary_1966} to answer this question.
Let $f = |\lambda|$ be the number of boxes in the Young diagram $\lambda$. As shown above $\lambda$ also labels a representation of the symmetric group on $f$ objects $S_f$. This representation is $\Delta_\lambda^f$. 
Take the Young diagram $\mu$ ($\nu$) and add columns of $m$ ($n$) boxes to the left until it has $f$ boxes.  The tableau obtained which we call $\mu_f$ ($\nu_f$) labels a representation of $S_f$ denoted $\Delta_{\mu_f}^f$ ($\Delta_{\nu_f}^f$).
We remember that adding columns to the left of length $m$ ($n$) keeps $\mu$ ($\nu$) within the equivalence class of Young diagrams labelling the representation $\Gamma_\mu^m$ ($\Gamma_\nu^n$). Hence $\mu_f$ ($\nu_f$) labels the same representation of $\mathrm{SU}(m)$ ($\mathrm{SU}(n)$) as $\mu$ ($\nu$).

Hence the diagrams $\lambda$ , $\mu_f$ and $\nu_f$ refer both to representations of the special unitary group  $\Gamma^{mn}_\lambda$ , $\Gamma^m_\mu (= \Gamma^m_{\mu_f} ) $ and  $\Gamma^n_\nu (= \Gamma^n_{\nu_f} )$ as well as representations of $S_f$:  $\Delta_\lambda^f$,  $\Delta_{\mu_f}^f$ and $\Delta_{\nu_f}^f$.

\begin{thm}\label{recipetheorem}
$\Gamma^m_\mu \boxtimes  \Gamma^n_\nu$ occurs as many times in the restriction of  $\Gamma^{mn}_\lambda$ to $ \mathrm{SU}(m) \times \mathrm{SU}(n)$ as $\Delta_\lambda^f$ occurs in $\Delta_{\mu_f}^f \otimes \Delta_{\nu_f}^f$, where $f = |\lambda|$~\cite{itzykson_unitary_1966}.
\end{thm}

\subsection{Inductive lemma}

\begin{lemma}\label{inductivelemma}
Consider representations $\Gamma^{mn}_{\bar \lambda}$, $\Gamma^{m}_{\bar \mu}$, $\Gamma^{n}_{\bar \nu}$, $\Gamma^{mn}_{ \lambda}$, $\Gamma^{m}_{ \mu}$, $\Gamma^{n}_{ \nu}$, $\Gamma^{mn}_{\lambda'}$, $\Gamma^{m}_{ \mu'}$ and $\Gamma^{n}_{ \nu'}$ where $\bar \lambda = \lambda + \lambda'$, $\bar \mu = \mu + \mu'$ $\bar \nu = \nu + \nu'$ and $\frac{|\lambda|-|\mu|}{m},\frac{|\lambda|-|\nu|}{n},\frac{|\lambda'|-|\mu'|}{m}$ and $\frac{|\lambda'|-|\nu'|}{n}$ are integers. If $\Gamma^{mn}_{ \lambda| \mathrm{SU}(m) \times \mathrm{SU}(n)}$ contains a term $\Gamma^{m}_{ \mu} \boxtimes \Gamma^{n}_{ \nu}$ and $\Gamma^{mn}_{ \lambda'| \mathrm{SU}(m) \times \mathrm{SU}(n)}$ contains a term $\Gamma^{m}_{ \mu'} \boxtimes \Gamma^{n}_{ \nu'}$ then $\Gamma^{mn}_{ \bar \lambda| \mathrm{SU}(m) \times \mathrm{SU}(n)}$ contains a term $\Gamma^{m}_{ \bar \mu} \boxtimes \Gamma^{n}_{ \bar \nu}$.
\end{lemma}

\begin{proof}
$\Gamma^{mn}_{ \lambda| \mathrm{SU}(m) \times \mathrm{SU}(n)}$ containing a term $\Gamma^{m}_{ \mu} \boxtimes \Gamma^{n}_{ \nu}$ implies that $\Delta^f_\lambda$ occurs in $\Delta^f_{\mu_f} \otimes \Delta^f_{\nu_f}$ (by Theorem \ref{recipetheorem}). Here $\mu_f$ is the tableau $\mu$ ($\nu$) to which  $\frac{f-|\mu|}{m}$($\frac{f-|\nu|}{n}$) columns of length $m$ ($n$) has been added so that the total number of boxes $|\mu_f| = f$ ($|\nu_f|=f$).

\begin{align}
& \mu_f = \mu + \left(\left(\frac{f-|\mu|}{m}\right)^m\right) \ , \\
& \nu_f = \nu + \left(\left(\frac{f-|\nu|}{n}\right) ^n\right)   \ .
\end{align}
Here we recall that $((\frac{f-|\mu|}{m})^m)$ indicates $m$ rows of length $((\frac{f-|\mu|}{m})$.
This implies that $g(\lambda, \mu_f, \nu_f)>0$. 
Similarly $g(\lambda', \mu'_{f'}, \nu'_{f'})>0$.

We now show that $\bar \mu = \mu + \mu'$ and $\bar \nu = \nu + \nu'$ implies that $\bar \mu_{\bar f} = \mu_f + \mu'_{f'}$ and $\bar \nu_{\bar f} = \nu_f + \nu'_{f'}$.

\begin{align}
\mu_f + \mu'_{f'} &  =  \mu + \mu' + \left(\left(\frac{f-|\mu|}{m}\right) ^m\right)  + \left(\left(\frac{f'-|\mu'|}{m}\right) ^m\right)   \\
& = \bar \mu + \left(\left(\frac{(f+f' - |\mu| -|\mu'|)}{m}\right) ^m\right)   \\
& =  \bar \mu + \left(\left(\frac{\bar f-|\bar \mu|}{m}\right) ^n\right)    = \bar \mu_{\bar f} \ .
\end{align}
And similarly for $\bar \nu$. 

Let us make use of a property of the Clebsch Gordan coefficients called the semi-group property. If $g(\lambda, \mu_f , \nu_f) > 0 $ and $g(\lambda', \mu'_{f'} ,\nu'_{f'}) > 0$ then $g(\lambda + \lambda',\mu_f + \mu'_{f'}  , \nu_f + \nu'_{f'} ) > 0$ \cite{ikenmeyer_rectangular_2016}. By the semi-group property we have $g(\bar \lambda, \bar \mu_{\bar f}, \bar \nu_{\bar f})>0$. This implies that $\Delta^{\bar f}_{
\bar \lambda}$ occurs in $\Delta^{\bar f}_{\bar \mu_{\bar f}} \otimes \Delta^{\bar f}_{\bar \mu_{\bar f}}$. By Theorem \ref{recipetheorem} this implies that $\Gamma^{mn}_{ \bar \lambda| \mathrm{SU}(m) \times \mathrm{SU}(n)}$ contains a term $\Gamma^{m}_{ \bar \mu} \boxtimes \Gamma^{m}_{ \bar \nu}$.
\end{proof}

\section{Violation of local tomography in all alternative measurement postulates}\label{nolocaltom}

In the following we establish that $\Gamma_{ | \suthree \times \suthree}^{9}$ is not of the form \eqref{localtomography} for all representations $\Gamma^{9}$ of $\sunine$ corresponding to non quantum state spaces with pure states $\mathbb C^9$. We show explicitly that every such restriction contains a term of $1^{3} \boxtimes 1^{3}$, where $1^{3}$ is the trivial representation of $\suthree$.

\subsection{Arbitrary dimension $d$}

We now construct a proof by induction to show that the representations $\mathcal D_j^d$ are not compatible with local tomography when restricted to ${\rm SU}(d_\sA) \times {\rm SU}(d_\sB)$. A representation $\mathcal D_j^d$ will violate local tomography if it is not of the form \eqref{localtomography} when restricted to ${\rm SU}(d_\sA) \times {\rm SU}(d_\sB)$. It suffices to show that there is a term $1^{d_\sA} \boxtimes 1^{d_\sB}$ in this restriction in order to show that it is not of this form.

\begin{lemma}\label{inductionlemma}
If the representations $\mathcal D_j^d$ and $\mathcal D_2^d$  of $\sud$ contain a term  $1^{d_\sA} \boxtimes 1^{d_\sB}$ when restricted to ${\rm SU}(d_\sA) \times {\rm SU}(d_\sB)$ then so does $\mathcal D_{j+2}^d$
\end{lemma}

\begin{proof}

Let

\begin{align}
& \Gamma_\lambda^d = \mathcal D_j^d ,  \ \Gamma_\mu^{d_\sA} = 1^{d_\sA}  , \  \Gamma_\nu^{d_\sB} = 1^{d_\sB}  \\
& \Gamma_{\lambda'}^{d} = \mathcal D_2^d , \ \Gamma_{\mu'}^{d_\sA}=1^{d_\sA}  , \ \Gamma_{\nu'}^{d_\sB} = 1^{d_\sB} \\
& \Gamma_{\bar \lambda} =  \mathcal D_{j+2}^{d} , \ \Gamma_{\bar \mu} = 1^{d_\sA}  , \ \Gamma_{\bar \nu}^{d_\sB} =  1^{d_\sB} \ ,
\end{align}

where

\begin{align}
& \lambda=(2j,j^{d-2}) , \ \mu = 0, \ \nu = 0 \ ,\\
& \lambda' = (4,2^{d-2}) , \ \mu' = 0 , \ \nu' = 0 \ , \\
& \bar \lambda= (2j+4,(j+2)^{d-2}) , \ \bar \mu = 0 , \ \bar \nu = 0  \ .
\end{align}

We observe

\begin{align}
& \bar \lambda = \lambda + \lambda' , \ \bar \mu = \mu + \mu' , \ \bar \nu = \nu + \nu' \\
& f = |\lambda| = jd , \ f' = |\lambda'| = 2d , \ \bar f = | \bar \lambda| = d(j+2)
\end{align}

Next we check that the quantities below are integer valued:

\begin{align}
\frac{|\lambda|-|\mu|}{m} = \frac{jd}{d_\sA} = j d_\sB \ ,\\
\frac{|\lambda|-|\nu|}{n}  = \frac{jd}{d_\sB} = j d_\sA \ , \\
\frac{|\lambda'|-|\mu'|}{m} = \frac{2d}{d_\sA} = 2d_\sB \ , \\ 
\frac{|\lambda'|-|\nu'|}{n} = \frac{2d}{d_\sB} =2d_\sA  \ .
\end{align}

From Lemma \ref{inductivelemma}  it follows that if $\mathcal D_j^d$ contains a representation $1^{d_\sA} \boxtimes 1^{d_\sB}$ and $\mathcal D_2^d$ contains a representation $1^{d_\sA} \boxtimes 1^{d_\sB}$ when restricted to ${\rm SU}(d_\sA) \times {\rm SU}(d_\sB)$ then $\mathcal D_{j+2}^d$ contains a representation  $1^{d_\sA} \boxtimes 1^{d_\sB}$ when restricted to ${\rm SU}(d_\sA) \times {\rm SU}(d_\sB)$.

\end{proof}

Hence it suffices to show that $\mathcal D_2^d$ and $\mathcal D_3^d$ contain $1^{d_\sA} \boxtimes 1^{d_\sB}$ when restricted to ${\rm SU}(d_\sA) \times {\rm SU}(d_\sB)$  to show that $\mathcal D_j^d$ does for any $j >1$.

\subsection{Existence of $1^3 \boxtimes 1^3$ for all non-quantum representations of $\sunine$}

Using Sage software~\cite{sagemath} we can show that  $\mathcal D_2^9$ and $\mathcal D_3^9$ have a representation $1^{3} \boxtimes 1^{3}$ when restricted to $\suthree \times \suthree$. By Lemma \ref{inductionlemma} all representations $\mathcal D_j^9$, $j>1$ have this property. An arbitrary representation corresponding to an alternative Born rule for $\pcnine$ is of the form:
\begin{equation}\label{nonquantnine}
\Gamma_9 =\bigoplus_{j \in \mathcal J}  \mathcal D_j^9
\end{equation}
Where $\mathcal J$ is a list of positive integers containing at least one integer which is not 1. Since at least one (non-trivial) subrepresentation in $\Gamma^9$ has a $1^{3} \boxtimes 1^{3}$ when restricted to $\suthree \times \suthree$ so does the representation $\Gamma^9$. 

\subsection{Violation of local tomography for all theories}

Every representation $\Gamma_9$ of $\sunine$ of the form \eqref{nonquantnine} has a representation $1^{3} \boxtimes 1^{3}$ when restricted to $\suthree \times \suthree$. It follows from this that the restriction of $\Gamma_9$ to $\suthree \times \suthree$ is not of the form required for local tomography. From this it follows that all non-quantum  $\pcnine$ systems which are composites of two $\pcthree$ systems violate local tomography.

In order to show that a theory with systems $\pcd$ (for every $d>1$) violates local tomography, it is sufficient to show that one of the systems in the theory violates local tomography. Since all $\pcnine$ non-quantum systems violate local tomography it follows that all non-quantum theories with systems $\pcd$ violate local tomography. We emphasise that here we consider theories for which all values of $d>1$ are possible.

\section{Composition in GPTs}

In the earlier appendices we gave a certain primacy to the space of OPFs, and considered the group action on the linear space spanned by the OPFs. However it is equally valid (and more common) to give primacy to the space of states and consider the group action on this space.
This standard representation will be helpful for proving the consistency of the toy model. Indeed the consistency constraints required for composition translate naturally to the language of states. 
This section is largely without proof, as the proofs carry over naturally from the OPF formalism to the standard state formalism. We refer the reader to \cite{Masanes_lecture_2017} for more detailed proofs.

\subsection{Representation of states}

The set of mixed states span a real vector space. A state is a vector:
\begin{equation}
\bar \omega  = \left(
\begin{array}{c}
F_1(\omega) \\
F_2(\omega)\\
\vdots \\
F_{K_d}(\omega)
\end{array}
\right) \ ,
\end{equation}
An arbitrary $F \in \mathcal F_d$ can be expressed:
\begin{equation}
F = \sum_i c_i F_i
\end{equation}
To each $F$ we can associate a dual vector (or effect) $\bar {\rm E}_F$
\begin{equation}
\bar {\rm E}_F = (c_1 , ... , c_{K_d}) \ ,
\end{equation}
By definition $\bar {\rm E}_F \cdot \bar \omega = F(\omega)$.
Let $\bar {\bf u}$ be effect associated to the unit OPF. Since $\bar {\bf u} \cdot \bar \omega = 1$ there is a component for all states $\bar \omega$ which is constant. That is to say we can choose $F_1$ to be the unit OPF:
\begin{equation}
\bar \omega  = \left(
\begin{array}{c}
1 \\
 \omega\\
\end{array}
\right) \ ,
\end{equation}
The group representation $ \Gamma^{d}$ acting on the state space is of the form $ \Gamma^d = 1^d +  \Gamma^d$ ($ \Gamma^d$ may be reducible or irreducible). 
We can write
\begin{equation}
\bar {\rm E}_F \cdot \bar \omega = c_1 +  {\rm E}_F \cdot  \omega 
\end{equation}

In the representation where the trivial component is removed (by an affine transformation) then states are $ \omega$ (whose fiducial outcomes are affinely independent), the group representation is $ \Gamma^d$ and outcome probabilities are affine functions $ {\rm E}_F$ of the state $\omega$.
We observe that from the uniqueness of the trivial component in $\Gamma^d$ it follows that $\Gamma^d$ cannot contain any trivial subrepresentation.

\subsection{Local tomography}

Consider measurement postulates $\mathcal F_{d_\sA d_\sB}$, $\mathcal F_{\sA}$ and $\mathcal F_{\sB}$. 
States for Alice and Bob can be written as:
\begin{equation}
\bar \omega_\sA  = \left(
\begin{array}{c}
F_\sA^1(\omega_\sA) \\
F_\sA^2(\omega_\sA)\\
\vdots \\
F_\sA^{K_{d_\sA}}(\omega_\sA)
\end{array}
\right) \ ,
\end{equation}
and
\begin{equation}
\bar \omega_\sB  = \left(
\begin{array}{c}
F_\sB^1(\omega_\sB) \\
F_\sB^2(\omega_\sB)\\
\vdots \\
F_\sB^{K_{d_\sB}}(\omega_\sB)
\end{array}
\right) \ .
\end{equation}

\begin{definition}[Local tomography]
	A composite system  $\mathcal S_{\sA\sB}$ is locally tomographic if it has fiducial outcomes $\{(F_\sA^i \star F_\sB^j) \}_{i = 1 , j = 1}^{i = K_{d_\sA} , j = K_{d_\sB}}$. A state of the composite system can be written:
	\begin{equation}
	\bar \omega_{\sA\sB}  = \left(
	\begin{array}{c}
	(F_\sA^1 \star F_\sB^1) (\omega_{\sA \sB}) \\
	(F_\sA^1 \star F_\sB^2) (\omega_{\sA \sB}) \\
	\vdots \\
	(F_\sA^{\kda} \star F_\sB^{\kdb}) (\omega_{\sA \sB}) 
	\end{array}
	\right) \ ,
	\end{equation}
\end{definition}

We observe that $\bar E_{F_\sA^i \star F_\sB^j} = \bar E_{F_\sA^i} \otimes \bar E_{F_\sB^j}$. Indeed any joint local effect $\bar E_{F_\sA \star F_\sB} = \bar E_{F_\sA} \otimes \bar E_{F_\sB}$. Product states are of the form $\bar \omega_\sA \otimes \bar \omega_\sB$.

\subsection{Non-locally-tomographic theories}

The states of a composite system $\mathcal S_{\sA\sB}$ of a non-locally tomographic (or holistic) theory can be written as~\cite{Hardy_foliable_2009,Masanes_lecture_2017}:
\begin{equation}
\bar \omega_{\sA\sB}  = \left(
\begin{array}{c}
\bar \lambda_{\sA\sB}\\
\eta_{\sA\sB}\\
\end{array}
\right) \ ,
\end{equation}
where
\begin{equation}
\bar \lambda_{\sA\sB}  = \left(
\begin{array}{c}
(F_\sA^1 \star F_\sB^1) (\omega_{\sA \sB}) \\
(F_\sA^1 \star F_\sB^2) (\omega_{\sA \sB}) \\
\vdots \\
(F_\sA^{\kda} \star F_\sB^{\kdb}) (\omega_{\sA \sB}) 
\end{array}
\right) \ ,
\end{equation}
is the locally tomographic part and 
\begin{equation}
\eta_{\sA\sB} = \left(
\begin{array}{c}
F_{\sA \sB}^1  (\omega_{\sA \sB})\\
\vdots\\
F_{\sA \sB}^{K_{d_\sA d_\sB} - K_{d_\sA} K_{d_\sB}} (\omega_{\sA \sB})
\end{array}
\right) \ ,
\end{equation}
is the holistic part. The probability for a joint outcome $F_\sA \star F_\sB$ is given by:
\begin{equation}\label{jointlocaleffect}
(F_\sA \star F_\sB) (\omega_{\sA \sB}) = (\bar E_{F_\sA} \otimes \bar E_{F_\sB}) \cdot \bar \lambda_{\sA \sB}
\end{equation}
Joint local effects are computed using the locally tomographic part of the state space only.

\subsection{The reduced state space}

The reduced states for Alice and Bob can be computed as follows:
\begin{align}
\bar \omega_\sA = (\bar \Gamma^{d_\sA}(\unity_\sA) \otimes    {\bf u}_\sB ) \bar \lambda_{\sA\sB} \label{reducedstate} \\
\bar \omega_\sB = ( {\bf u}_\sA \otimes \bar \Gamma^{{d_\sB}}(\unity_\sB)  ) \bar \lambda_{\sA\sB}
\end{align}
Reduced states are determined using only the locally tomographic part of the global state.

\section{Toy model}\label{toymodel}

In this appendix we show that the toy model introduced in section~\ref{Toymodel} meets consistency constraints {\bf C1 - C5} (apart from associativity of the $\star$ product). 

\subsection{Constraints C1 and  C2}

It is immediate that consistency constraints {\bf C1} and {\bf C2} are met by the toy model.

\subsection{Constraint C3}

We prove $  (F_\sA \star F_\sB)   (\psi_\sA \otimes \phi_\sB) = F_\sA (\psi_\sA) F_\sB (\phi_\sB)\ $: 

\begin{align}
& (F_\sA \star F_\sB)   (\psi_\sA \otimes \phi_\sB) \nonumber \\ 
= &  \  {\rm tr}\Big( \ketbra{\psi}{\psi}_\sA^{\otimes 2}\ketbra{\phi}{\phi}_\sB^{\otimes 2}(\hat F_\sA \hat  F_\sB     + \frac{{\rm tr} (\hat F_\sA \hat  F_\sB )}{{\rm tr}(S_\sA S_\sB)}A_\sA A_\sB)\Big) \nonumber \\
 =  &     {\rm tr}\Big( \ketbra{\psi}{\psi}_\sA^{\otimes 2}\ketbra{\phi}{\phi}_\sB^{\otimes 2} \hat F_\sA \hat  F_\sB \Big)  \nonumber \\
= & \  F_\sA (\psi_\sA) F_\sB (\phi_\sB) \ .
\end{align}
In the penultimate line we have used the fact that the overlap of product states $\ketbra{\psi}{\psi}_\sA^{\otimes 2}\ketbra{\phi}{\phi}_\sB^{\otimes 2}$ and $A_\sA A_\sB$ is $0$.

\subsection{Constraint C4}

In the following it will occasionally useful to label the two copies of $\mathbb C^{d_\sA}$ with $1$ and $3$ and to label the two copies of $\mathbb C^{d_\sB}$ with $2$ and $4$. We write $\tilde S_\sA$ for the normalised projector onto the symmetric subspace of $(\mathbb C^{d_\sA})^{\otimes 2}$. We make use of the identity $S = \frac{1}{2}(\unity + {\rm SWAP})$ throughout this section.

In this section we show that normalised conditional states for Alice are valid states of a $\mathbb C^{d_\sA}$ system. 
We first show this for the specific case where the state is conditioned on the unit effect, i.e. is a reduced state. A reduced state of Alice for a bi-partite system in pure state $\ketbra{\psi_{\sA\sB}}{\psi_{\sA\sB}}^{\otimes 2}$ is:
\begin{align}
\bar \omega_\sA 
=
 {\rm tr}_\sB \! \left( S_\sB \ketbra{\psi_{\sA\sB}}{\psi_{\sA\sB}}^{\otimes 2} \right) 
+ \frac {S_\sA}{{\rm tr} S_\sA} 
{\rm tr}_{\sA\sB} \! 
\left(A_\sA  A_\sB 
\ketbra{\psi_{\sA\sB}}{\psi_{\sA\sB}}^{\otimes 2} \right) \	.
\end{align}

We show that these reduced states lie in the convex hull of $\ketbra{\psi_\sA}{\psi_\sA}^{\otimes 2}$. 

\begin{lemma}\label{simpequivlem}
	$S_\sB \ket{\psi_{\sA\sB}}^{\otimes 2}  = S_\sA S_\sB \ket{\psi_{\sA\sB}}^{\otimes 2}$
\end{lemma}

\begin{proof}
	\begin{equation}
	\ket{\psi_{\sA\sB}}^{\otimes 2} = \alpha_{i_1 i_2} \alpha_{i_3 i_4} \ket{i_1 i_2 i_3 i_4} \ .
	\end{equation}
	\begin{equation}
	S_\sB \ket{\psi_{\sA\sB}}^{\otimes 2} = \frac{1}{2} \alpha_{i_1 i_2} \alpha_{i_3 i_4} (\ket{i_1 i_2 i_3 i_4} +\ket{i_1 i_4 i_3 i_2}) \ .
	\end{equation}
	Let us relabel $i_1 \leftrightarrow i_3$ in the last term:
	\begin{align}
	S_\sB \ket{\psi_{\sA\sB}}^{\otimes 2} = \frac{1}{2} ( \alpha_{i_1 i_2} \alpha_{i_3 i_4} \ket{i_1 i_2 i_3 i_4} + \alpha_{i_3 i_2} \alpha_{i_1 i_4} \ket{i_3 i_4 i_1 i_2}) \ .
	\end{align}
	\begin{equation}
	S_\sA S_\sB \ket{\psi_{\sA\sB}}^{\otimes 2} = \frac{1}{4} \alpha_{i_1 i_2} \alpha_{i_3 i_4} (\ket{i_1 i_2 i_3 i_4} +\ket{i_1 i_4 i_3 i_2}+\ket{i_3 i_2 i_1 i_4} +\ket{i_3 i_4 i_1 i_2}) \ .
	\end{equation}
	Let us relabel $i_1 \leftrightarrow i_3$ in the second term , $i_2 \leftrightarrow i_4$ in the penultimate term and  $i_1 \leftrightarrow i_3$ and $i_2 \leftrightarrow i_4$ in the last term :
	\begin{align}
	& S_\sA S_\sB \ket{\psi_{\sA\sB}}^{\otimes 2}  
	=  \frac{1}{4} (\alpha_{i_1 i_2} \alpha_{i_3 i_4} \ket{i_1 i_2 i_3 i_4}  +\alpha_{i_3 i_2} \alpha_{i_1 i_4}  \ket{i_3 i_4 i_1 i_2} \\
	+ &  \alpha_{i_1 i_4} \alpha_{i_3 i_2} \ket{i_3 i_4 i_1 i_2} + \alpha_{i_1 i_2} \alpha_{i_3 i_4} \ket{i_1 i_2 i_3 i_4}) 
	=  \frac{1}{2} ( \alpha_{i_1 i_2} \alpha_{i_3 i_4} \ket{i_1 i_2 i_3 i_4} + \alpha_{i_3 i_2} \alpha_{i_1 i_4} \ket{i_3 i_4 i_1 i_2}) \ .
	\end{align}
\end{proof}

From the above lemma we can write a reduced state $\bar \omega$:
\begin{equation}
\bar \omega_\sA 
=
{\rm tr}_\sB \! \left( S_\sA S_\sB \ketbra{\psi_{\sA\sB}}{\psi_{\sA\sB}}^{\otimes 2} \right)
+ \frac {S_\sA}{{\rm tr} S_\sA} 
{\rm tr}_{\sA\sB} \! 
\left(A_\sA  A_\sB 
\ketbra{\psi_{\sA\sB}}{\psi_{\sA\sB}}^{\otimes 2} \right) \ .
\end{equation}

\begin{lemma}\label{Redstatelem}
	The reduced state $\bar \omega_\sA$ can be written as:
	\begin{equation}
	\bar \omega_\sA =  S_\sA(\rho_\sA \otimes \rho_\sA) S_\sA + (1 - {\rm tr}( S_\sA(\rho_\sA \otimes \rho_\sA) S_\sA)) \tilde S_\sA \ ,
	\end{equation}
	where $\rho_\sA = {\rm tr}_\sB (\ketbra{\psi}{\psi}_{\sA \sB})$.
\end{lemma}

\begin{proof}
	We first show that:
	\begin{equation}
	{\rm tr}_\sB ( S_\sA S_\sB \ketbra{\psi}{\psi}^{\otimes 2}_{\sA \sB}) = S_\sA (\rho_\sA \otimes \rho_\sA) S_\sA \ .
	\end{equation}
	From the proof of Lemma ~\ref{simpequivlem}:
	\begin{align}
	S_\sA S_\sB \ket{\psi_{\sA\sB}}^{\otimes 2} = \frac{1}{4} \alpha_{i_1 i_2} \alpha_{i_3 i_4} (\ket{i_1 i_2 i_3 i_4} +\ket{i_1 i_4 i_3 i_2}+\ket{i_3 i_2 i_1 i_4} +\ket{i_3 i_4 i_1 i_2}) \ .
	\end{align}
	Hence:
	\begin{align}
	S_\sA S_\sB \ketbra{\psi_{\sA\sB}}{\psi_{\sA\sB}}^{\otimes 2} = & \frac{1}{4} \alpha_{i_1 i_2} \alpha_{i_3 i_4}  \bar \alpha_{j_1 b_1}  \bar \alpha_{j_3 b_2} (\ketbra{i_1 i_2 i_3 i_4}{j_1 j_2 j_3 j_4}  \\
	+ & \ketbra{i_1 i_4 i_3 i_2}{j_1 j_2 j_3 j_4}+\ketbra{i_3 i_2 i_1 i_4}{j_1 j_2 j_3 j_4} +\ketbra{i_3 i_4 i_1 i_2}{j_1 j_2 j_3 j_4}) \ .
	\end{align}
	which implies:
	\begin{align}
	{\rm tr}_\sB (S_\sA  S_\sB \ketbra{\psi}{\psi}_{\sA\sB}^{\otimes 2}) 
	=  \frac{1}{2} \alpha_{i_1 b_1}  \alpha_{i_3 b_2}  \bar \alpha_{j_1 b_1}  \bar \alpha_{j_3 b_2} ( \ketbra{i_1  i_3 }{j_1  j_3 } +  \ketbra{i_3  i_1 }{j_1  j_3} ) \ .
	\end{align}
	We now compute $ S_\sA ( \rho_\sA \otimes \rho_\sA) S_\sA$.
	\begin{equation}
	\ketbra{\psi}{\psi}_{\sA \sB} = \alpha_{i_1 i_2} \bar \alpha_{j_1  j_2} \ketbra{i_1 i_2}{j_1  j_2} \ .
	\end{equation}
	\begin{equation}
	\rho_\sA = {\rm tr}_{\sB} (\ketbra{\psi}{\psi}_{\sA \sB}) = \alpha_{i_1 b_1} \bar \alpha_{j_1  b_1} \ketbra{i_1 }{j_1 } \ .
	\end{equation}
	\begin{equation}
	\rho_\sA \otimes \rho_\sA = \alpha_{i_1 b_1} \alpha_{i_3 b_2} \bar \alpha_{j_1  b_1} \bar \alpha_{j_3  b_2} \ketbra{i_1 i_3}{j_1  j_3} \ .
	\end{equation}
	\begin{align}
	S_\sA (\rho_\sA \otimes \rho_\sA ) 
	=  \frac{1}{2} \alpha_{i_1 b_1} \alpha_{i_3 b_2} \bar \alpha_{j_1  b_1} \bar \alpha_{j_3  b_2} (\ketbra{i_1 i_3}{j_1  j_3} +  \ketbra{i_3 i_1}{j_1  j_3}) \ .
	\end{align}
	\begin{align}
	S_\sA (\rho_\sA \otimes \rho_\sA ) S_\sA 
	=  & \frac{1}{4} \alpha_{i_1 b_1} \alpha_{i_3 b_2} \bar \alpha_{j_1  b_1} \bar \alpha_{j_3  b_2} (\ketbra{i_1 i_3}{j_1  j_3} +  \ketbra{i_3 i_1}{j_1  j_3})  \\
	+  & \frac{1}{4} \alpha_{i_1 b_1} \alpha_{i_3 b_2} \bar \alpha_{j_1  b_1} \bar \alpha_{j_3  b_2} (\ketbra{i_1 i_3}{j_3  j_1} +  \ketbra{i_3 i_1}{j_3  j_1}) \ .
	\end{align} 
	In the second term we relabel $j_1 \leftrightarrow j_3$, $i_1 \leftrightarrow i_3$ and $b_1 \leftrightarrow b_2$ to obtain:
	\begin{align}
	S_\sA (\rho_\sA \otimes \rho_\sA ) S_\sA  = \frac{1}{2} (\alpha_{i_1 b_1}  \alpha_{i_3 b_2}  \bar \alpha_{j_1 b_1}  \bar \alpha_{j_3 b_2} ( \ketbra{i_1  i_3 }{j_1  j_3 } +  \ketbra{i_3  i_1 }{j_1  j_3} ) ) \ .
	\end{align}
	Hence,
	\begin{align}
	\bar \omega_\sA & = S_\sA (\rho_\sA \otimes \rho_\sA) S_\sA + (1-{\rm tr} (\ketbra{\psi}{\psi}^{\otimes 2}_{\sA \sB} S_\sA S_\sB)) \tilde S_\sA \nonumber \\
	& = S_\sA(\rho_\sA \otimes \rho_\sA) S_\sA + (1 - {\rm tr}( S_\sA(\rho_\sA \otimes \rho_\sA) S_\sA)) \tilde S_\sA \ .
	\end{align}
\end{proof}
\begin{lemma}\label{RedstatesLem}
	The reduced states $\bar \omega_\sA$ belong to the convex hull of the local pure states $\ketbra{\psi}{\psi}^{\otimes 2}$.
\end{lemma}
\begin{proof}
	By Lemma \ref{Redstatelem} the reduced state can be written as:
	\begin{equation}
	\bar \omega_\sA =  S_\sA(\rho_\sA \otimes \rho_\sA) S_\sA + (1 - {\rm tr}( S_\sA(\rho_\sA \otimes \rho_\sA) S_\sA)) \tilde S_\sA \ ,
	\end{equation}
	where $\rho_\sA = {\rm tr}_\sB (\ketbra{\psi}{\psi}_{\sA \sB})$. In the following we drop the $\sA$ label. 
	\begin{equation}
	\rho = \sum_i \alpha_i \ketbra{i}{i} \ ,
	\end{equation}
	Here the $\ket i$ are not necessarily orthogonal. The trace of $\rho$ is $\sum_i \alpha_i = 1$, where $\alpha_i >0$.
	Let us write $\Phi_{ij} = \frac{1}{\sqrt{2}} (\ket{i,j} + \ket{j, i})$ and observe:
	\begin{equation}
	\sum_{i \neq j} \ketbra{\Phi_{ij}}{\Phi_{ij}} =  2 \sum_{i<j}  \ketbra{\Phi_{ij}}{\Phi_{ij}} = \sum_{i \neq j} \left(\ketbra{i,j}{i,j} + \ketbra{i,j}{j,i} \right)  \ .
	\end{equation}

	Consider the (not necessarily normalised) matrix
	\begin{equation}
	\label{projection}
	S (\rho \otimes \rho) S = S(\sum_{ij} \alpha_{i} \alpha_j \ketbra{i,j}{i,j})S 
	=
	\sum_{i} \alpha_i^2 |i,i\rangle\! \langle i,i|
	+ \sum_{i<j} \alpha_i \alpha_j \ketbra{\Phi_{ij}}{\Phi_{ij}}\ ,
	\end{equation}
	
	The trace of this matrix is $1- \sum_{i<j} \alpha_i \alpha_j$; hence:
	\begin{equation}
	\label{target}
	\bar \omega =   \sum_{i} \alpha_i^2 |i,i\rangle\! \langle i,i|
	+ \sum_{i<j} \alpha_i \alpha_j \ketbra{\Phi_{ij}}{\Phi_{ij}}
	+ \sum_{i<j} \alpha_i \alpha_j \tilde S \ .
	\end{equation}
	We now show that this arbitrary mixed state $\bar \omega$ can be expressed as a convex combination of local pure states $\ketbra{\psi}{\psi}^{\otimes 2}$.
	Consider the general vector
	\begin{equation}
	|\psi\rangle =
	\sum_i e^{i \theta_j} \sqrt{\alpha_j} |j\rangle\ ,
	\end{equation}
	where $\alpha_j \geq 0$ and  for all $j$. Normalisation implies $\sum_j \alpha_j =1$.
	Now, let us write the pure product state
	\begin{equation}
	\label{sim prod}
	|\psi\rangle\! \langle\psi|^{\otimes 2} =
	\sum_{j,k,j',k'} 
	e^{i \theta_j} e^{i \theta_k} e^{-i \theta_{j'}}  e^{-i \theta_{k'}}
	\sqrt{\alpha_j \alpha_k \alpha_{j'} \alpha_{k'}}\,
	|j,k\rangle\! \langle j',k'|\ .
	\end{equation}
	Let us make the following observations. When $j \neq j'$:
	\begin{equation}
	\int_{- \pi}^{\pi} e^{ - i \theta_j}  e^{i \theta_{j'}} d \theta_j d \theta_{j'} = 0 \ .
	\end{equation}
	When $j = j'$:
	\begin{equation}
	\int_{- \pi}^{\pi} e^{ - i \theta_j}  e^{i \theta_{j'}} d \theta_{j} d \theta_{j'} = (2 \pi)^2 \ .
	\end{equation}
	Now consider:
	\begin{align}
	\mathbb E_{\theta_i }
	|\psi\rangle\! \langle\psi|^{\otimes 2}  = 
	\frac{1}{(2 \pi)^4}\int_{-\pi}^{\pi}  \sum_{j,k,j',k'} 
	e^{i \theta_j} e^{i \theta_k} e^{-i \theta_{j'}}  e^{-i \theta_{k'}}
	\sqrt{\alpha_j \alpha_k \alpha_{j'} \alpha_{k'}}\,
	|j,k\rangle\! \langle j',k'| d \theta_j d \theta_k d \theta_{j'}  d \theta_{k'} \ .
	\end{align}
	The non zero contributions will arise from the following terms:

	\noindent
	$j=j'=k=k'$:
	\begin{equation}\label{phase1}
	\int_{- \pi}^{\pi}|e^{i \theta_j}|^4 d \theta_j d \theta_k d \theta_{j'}  d \theta_{k'}= (2\pi)^4 \ .
	\end{equation}
	
	\noindent
	$j = j' \neq k = k'$: 
	\begin{equation}\label{phase2}
	\int_{- \pi}^{\pi}|e^{i \theta_j}|^2 |e^{i \theta_k}|^2 d \theta_j d \theta_k d \theta_{j'}  d \theta_{k'}= (2\pi)^4 \ .
	\end{equation}
	
	\noindent
	$j = k' \neq k = j'$: 
	\begin{equation}\label{phase3}
	\int_{- \pi}^{\pi}|e^{i \theta_j}|^2 |e^{i \theta_k}|^2 d \theta_j d \theta_k d \theta_{j'}  d \theta_{k'}= (2\pi)^4 \ .
	\end{equation}
	All other contributions will be zero.

	Now, we write the mixed state corresponding to the uniform average over all values of the phases $\theta_i$,
	\begin{equation}
	\bar \omega_1 =   \mathbb E_{\theta_i }
	|\psi\rangle\! \langle\psi|^{\otimes 2} 
	=
	\sum_{j,k} 
	\alpha_j \alpha_k 
	|j,k\rangle\! \langle j,k| 
	+ \sum_{j\neq k} \alpha_j \alpha_k  |j,k\rangle\! \langle k,j| \ ,
	\end{equation}
	where the first term arises from contributions ~\eqref{phase1} and~\eqref{phase2} and the second term arises from contribution~\eqref{phase3}. Then we can write
	\begin{equation}
	\bar \omega_1
	=
	\sum_{i} \alpha_i^2 |i,i\rangle\! \langle i,i|
	+ 2 \sum_{i<j} \alpha_i \alpha_j \ketbra{\Phi_{ij}}{\Phi_{ij}} \ ,
	\end{equation}
	Let us take the state:
	\begin{equation}  
	\bar \omega_2 = \sum_{i} \alpha_i^2 |i,i\rangle\! \langle i,i| + 2 \sum_{i<j} \alpha_i \alpha_j \tilde S \ .
	\end{equation}
	This is a mixture of states of the form $\ketbra{\psi}{\psi}^{\otimes 2}$ since $\tilde S = \int \ketbra{\psi}{\psi}^{\otimes 2}d\psi$ and $\sum_{i} \alpha_i^2 + 2 \sum_{i<j} \alpha_i \alpha_j = 1$. 
	If we take the mixture $\frac{1}{2}(\bar \omega_1 + \bar \omega_2)$ we obtain~(\ref{target}).
\end{proof}
We now consider the more general case where Alice's state is conditioned on an arbitrary effect $F_\sB$.
The conditional state for Alice given one of Bob's effects $ F_\sB$ is:
\begin{align}
\bar \omega_{\sA | F_\sB} & = {\rm Tr}_\sB(S_\sA \hat  F_\sB \ketbra{\psi}{\psi}_{\sA \sB}^{\otimes 2} ) + {\rm Tr} (\ketbra{\psi}{\psi}^{\otimes 2}_{\sA \sB} A_\sA A_\sB)\frac{{\rm Tr}(\hat F_\sB)}{{\rm Tr}(S_\sB)} \tilde S_\sA  \ .
\end{align}
Although effects of the form $\hat F_\sB = \ketbra{\phi}{\phi}^{\otimes 2}$ are not valid (since the complement effects would not be of the required form), we calculate the conditional state for such effects, as this will allow us to later determine conditional states for general effects $\hat F_\sB = \sum_i \alpha_i \ketbra{\phi_i}{\phi_i}^{\otimes 2}$.
\begin{lemma}\label{equlemmacond}
	For $\hat F_\sB = \ketbra{\phi}{\phi}^{\otimes 2}$: 
	\begin{equation}
	{\rm Tr}_\sB(S_\sA \hat F_\sB \ketbra{\psi}{\psi}_{\sA \sB}^{\otimes 2} ) = S_\sA (\rho_{\sA |\phi}^\psi \otimes \rho_{\sA | \phi}^\psi) S_\sA = \rho_{\sA |\phi}^\psi \otimes \rho_{\sA |\phi}^\psi \ ,
	\end{equation}
	where $\rho_{\sA |\phi}^\psi = {\rm Tr}((\unity_\sA \otimes \ketbra{\phi}{\phi}_\sB)  \ketbra{\psi}{\psi}_{\sA \sB})$ \ .
\end{lemma}

\begin{proof}
	\begin{align}
	{\rm Tr}_\sB(S_\sA \hat F_\sB \ketbra{\psi}{\psi}_{\sA \sB}^{\otimes 2} ) = S_\sA {\rm Tr}_\sB( \hat F_\sB \ketbra{\psi}{\psi}_{\sA \sB}^{\otimes 2} ) = S_\sA (\rho_{\sA | \phi}^\psi \otimes \rho_{\sA | \phi}^\psi)  \ .
	\end{align}
	Let
	\begin{align}
	\rho_{\sA | \phi}^\psi = {\rm Tr} \big((\unity_\sA \otimes \ketbra{\phi}{\phi}_\sB) \ketbra{\psi}{\psi}_{\sA \sB} \big) 	= \sum_{i_1, j_1} \alpha_{i_1 , \phi} \bar \alpha_{j_1 , \phi} \ketbra{i_1}{j_1} \ .
	\end{align}
	We assume without loss of generality that $\ket \phi$ is one of the basis vectors $\ket i$.
	\begin{align}
	\rho_{\sA | \phi}^\psi  \otimes \rho_{\sA | \phi}^\psi = \sum_{i_1, i_3, j_1, j_3} \alpha_{i_1 , \phi} \bar \alpha_{j_1 , \phi} \alpha_{i_3 , \phi} \bar \alpha_{j_3 , \phi}  \ketbra{i_1 i_3}{j_1 j_3}  \ .
	\end{align}
	\begin{align}
	S_{\sA} (\rho_{\sA |  \phi}  \otimes \rho_{\sA |  \phi}) 
	=  & \frac{1}{2} (\sum_{i_1, i_3, j_1, j_3} \alpha_{i_1 ,  \phi} \bar \alpha_{j_1 , \phi} \alpha_{i_3 ,  \phi} \bar \alpha_{j_3 ,  \phi}  \ketbra{i_1 i_3}{j_1 j_3} \nonumber \\
	+  & \sum_{i_1, i_3, j_1, j_3} \alpha_{i_1 , \phi} \bar \alpha_{j_1 ,  \phi} \alpha_{i_3 ,  \phi} \bar \alpha_{j_3 ,  \phi}  \ketbra{i_3 i_1}{j_1 j_3} ) \ .	
	\end{align}
	We relabel $i_1 \leftrightarrow i_3$ on the last line to obtain $ S_\sA (\rho_{\sA | \phi} \otimes \rho_{\sA |  \phi})  = (\rho_{\sA |  \phi} \otimes \rho_{\sA |  \phi})$.
	\begin{align}
	& S_{\sA} (\rho_{\sA |  \phi}  \otimes \rho_{\sA |  \phi}) S_{\sA} 
	= \frac{1}{4} (\sum_{i_1, i_3, j_1, j_3} \alpha_{i_1 ,  \phi} \bar \alpha_{j_1 , \phi} \alpha_{i_3 ,  \phi} \bar \alpha_{j_3 ,  \phi}  \ketbra{i_1 i_3}{j_1 j_3} \nonumber \\
	+ & \sum_{i_1, i_3, j_1, j_3} \alpha_{i_1 , \phi} \bar \alpha_{j_1 ,  \phi} \alpha_{i_3 ,  \phi} \bar \alpha_{j_3 ,  \phi}  \ketbra{i_3 i_1}{j_1 j_3} 	+ \sum_{i_1, i_3, j_1, j_3} \alpha_{i_1 ,  \phi} \bar \alpha_{j_1 , \phi} \alpha_{i_3 ,  \phi} \bar \alpha_{j_3 ,  \phi}  \ketbra{i_1 i_3}{j_3 j_1} \nonumber \\
	+ & \sum_{i_1, i_3, j_1, j_3} \alpha_{i_1 , \phi} \bar \alpha_{j_1 ,  \phi} \alpha_{i_3 ,  \phi} \bar \alpha_{j_3 ,  \phi}  \ketbra{i_3 i_1}{j_3 j_1} ) \ .
	\end{align}
	We relabel $j_1 \leftrightarrow j_3$ on the last two lines to obtain $ S_\sA (\rho_{\sA | \phi} \otimes \rho_{\sA |  \phi})  = S_\sA (\rho_{\sA |  \phi} \otimes \rho_{\sA |  \phi})S_\sA $.
\end{proof}

We need one more lemma before proving that normalised conditional states  belong to the convex hull of the local pure states $\ketbra{\psi}{\psi}_\sA^{\otimes 2}$.

\begin{lemma}\label{stateform}
	If $S(\rho \otimes \rho) S = \rho \otimes \rho$ with ${\rm Tr} (\rho) = 1$ then $\rho \otimes \rho = \sum_i p_i \ketbra{\psi_i}{\psi_i}^{\otimes 2}$.
\end{lemma}

\begin{proof}
	By Lemma \ref{RedstatesLem} this is a valid reduced state and belongs to ${\rm conv} (\ketbra{\psi}{\psi}^\otimes 2)$.
\end{proof}

We observe that since all pure states are such that $S (\ketbra{\psi}{\psi}^{\otimes 2})S = \ketbra{\psi}{\psi}^{\otimes 2}$ and ${\rm Tr}(\ketbra{\psi}{\psi}^{\otimes 2}) = 1$, we can characterise the state space of the systems of the toy model as being given by ${\rm conv} (\rho \otimes \rho)$ for all normalised density operators such that $S(\rho \otimes \rho) S = \rho \otimes \rho$.

\begin{lemma}\label{CondstatesLem}
	The normalised conditional states $\tilde \omega_{\sA|F_\sB}$ belong to the convex hull of the local pure states $\ketbra{\psi}{\psi}^{\otimes 2}$.
\end{lemma}

\begin{proof}
	We first show that the conditional state is a valid local state for effects $\hat F_\sB = \ketbra{\phi}{\phi}^{\otimes2}$.
	As a conditional state, this state can be subnormalised. 
	\begin{align}
	\bar \omega_{\sA | F_\sB} 	 = {\rm Tr}_\sB(S_\sA \hat  F_\sB \ketbra{\psi}{\psi}_{\sA \sB}^{\otimes 2} ) + c \tilde S_\sA  \ ,
	\end{align}
	where $c = {\rm Tr} (\ketbra{\psi}{\psi}^{\otimes 2}_{\sA \sB} A_\sA A_\sB)\frac{{\rm Tr}(\hat F_\sB)}{{\rm Tr}(S_\sB)}$. 
	By Lemma \ref{equlemmacond} we have the equivalence:
	\begin{equation}
	{\rm Tr}_\sB(S_\sA \hat F_\sB \ketbra{\psi}{\psi}_{\sA \sB}^{\otimes 2} ) = (\rho_{\sA | \phi}^\psi \otimes \rho_{\sA | \phi}^\psi)  \ .
	\end{equation}
	The normalised conditional state is:
	\begin{equation}
	\tilde \omega_{\sA | F_\sB} = \frac{\bar \omega_{\sA | F_\sB}}{{\rm Tr}(\bar \omega_{\sA | F_\sB})} \ .
	\end{equation}
	Let us set $e= {\rm Tr}(\bar \omega_{\sA | F_\sB})$ and $d = {\rm Tr}(S_\sA \hat  F_\sB \ketbra{\psi}{\psi}_{\sA \sB}^{\otimes 2} ) ={\rm Tr} (\rho_{\sA | \phi}^\psi \otimes \rho_{\sA | \phi}^\psi)  $; $e = c+d$. 
	\begin{align}
	\tilde \omega_{\sA | F_\sB} 	=   \frac{1}{e}  (\rho_{\sA | \phi} \otimes \rho_{\sA | \phi})  
	+ \frac{c}{e}  \tilde S_\sA \ .
	\end{align}
	We use the equality $\rho_{\sA |\phi}^\psi \otimes \rho_{\sA | \phi}^\psi = d (\tilde \rho_{\sA | \phi}^\psi \otimes \tilde \rho_{\sA | \phi}^\psi) $ where $\tilde \rho_{\sA | F}^\psi$ is a standard normalised quantum conditional state.
	\begin{align}
	\tilde \omega_{\sA | F_\sB} = & \frac{d}{e}(\tilde \rho_{\sA | \phi}^\psi \otimes \tilde \rho_{\sA |\phi}^\psi) + \frac{c}{e} \tilde S_\sA  \ .
	\end{align}
	By Lemma \ref{equlemmacond} $\tilde \rho_{\sA | \phi}^\psi \otimes \tilde \rho_{\sA |\phi}^\psi = S_\sA (\tilde \rho_{\sA | \phi}^\psi \otimes \tilde \rho_{\sA |\phi}^\psi) S_\sA$ hence by Lemma \ref{stateform} it is a valid normalised state (i.e. of the form $\sum_i p_i \ketbra{\psi_i}{\psi_i}^{\otimes 2}$).
	Since $0 <\frac{d}{e} < 1$, $0 < \frac{c}{e} < 1 $ and $\frac{d + c}{e} = 1$ the above is a convex combination of $\tilde \rho_{\sA | \phi}^\psi \otimes \tilde \rho_{\sA |\phi}^\psi$ and $ \tilde S_\sA $ which are both valid states. Hence the state $\tilde \omega_{\sA | F_\sB}$ is a valid local state.
	
	Let $F_\sB = \sum \alpha_i \ketbra{\phi_i}{\phi_i}^{\otimes 2}$, with $\alpha_i > 0$. 
	
	\begin{align}
	\bar \omega_{\sA | F_\sB}  & =  \sum_{i} \alpha_i ({\rm Tr}_\sB(S_\sA \hat  F_\sB \ketbra{\phi_i}{\phi_i}^{\otimes 2}\ketbra{\psi}{\psi}_{\sA \sB}^{\otimes 2} ) 
	+ {\rm Tr} (\ketbra{\psi}{\psi}^{\otimes 2}_{\sA \sB} A_\sA A_\sB)\frac{1}{{\rm Tr}(S_\sB)} \tilde S_\sA) \nonumber \\
	& = \sum_{i} \alpha_i \big( (\rho_{\sA |\phi_i}^{\psi} \otimes \rho_{\sA |\phi_i}^{\psi}) + c_i \tilde S_\sA)  \ . 
	\end{align}
	Let $e = {\rm tr} (\omega_{\sA | F_\sB})$ and $d_i = {\rm tr}( \rho_{\sA |\phi_i}^{\psi} \otimes \rho_{\sA |\phi_i}^{\psi})$. We have $e = \sum_i \alpha_i (c_i + d_i)$.  From above:
	\begin{align}
	\tilde \omega_{\sA | F_\sB} = & \sum_i \alpha_i(  \frac{d_i}{e}(\tilde \rho_{\sA | \phi_i}^{\psi} \otimes \tilde \rho_{\sA | \phi_i}^{\psi}) + \frac{c_i}{e} \tilde S_\sA) \ .
	\end{align} 
	Since $0 < \frac{\alpha_ i d_i}{e} < 1$ and $0 < \sum_i \frac{ \alpha_ i c_i}{e}<1$ and $\sum_i \frac{  \alpha_i (c_i + d_i)}{e} = 1$ the above is a convex combination of $(\tilde \rho_{\sA | \phi_i}^{\psi} \otimes \tilde \rho_{\sA | \phi_i}^{\psi})$ with coefficients $\frac{\alpha_ i d_i}{e}$ and the state $\tilde S_\sA$ with coefficient $\sum_i \frac{\alpha_i c_i}{e}$.  
\end{proof}

\subsection{Constraint C5}

In this section we show that every OPFs in $\F_{d_\sA d_\sB}^\sg$ applied to a product state $\psi_\sA \otimes \phi_\sB$ has a corresponding OPF $F_\sA'$ in $\F_{d_\sA}^\sl$.
Let $F_{\sA \sB}$ be an arbitrary effect in $\F_{d_\sA d_\sB}^\sl$. The corresponding operator is
\begin{equation}
\hat F_{\sA \sB} = \sum_i \alpha_i \ketbra{x_i}{x_i}_{12} \otimes \ketbra{x_i}{x_i}_{34} \ .
\end{equation}
We evaluate it on product states:
\begin{align}
F_{\sA \sB} (\psi_\sA \otimes \phi_\sB) 
&  = \sum_i \alpha_i {\rm Tr}\big(\ketbra{\psi}{\psi}_\sA^{\otimes 2} \ketbra{\phi}{\phi}_\sB^{\otimes 2} \ (\ketbra{x_i}{x_i}_{\sA \sB}^{\otimes 2})\big) \nonumber \\
& = \sum_i \alpha_i {\rm Tr}_{\sA} (\ketbra{\psi}{\psi}_\sA^{\otimes 2} {\rm Tr}_{\sB} ( \ketbra{\phi}{\phi}_\sB^{\otimes 2} (\ketbra{x_i}{x_i}_{\sA \sB}^{\otimes 2}))  ) \nonumber \\
& = \sum_i \alpha_i  {\rm Tr}_{\sA} (\ketbra{\psi}{\psi}_\sA^{\otimes 2} S_\sA {\rm Tr}_{\sB}( \unity_\sA \ketbra{\phi}{\phi}_\sB^{\otimes 2}\ketbra{x_i}{x_i}_{\sA \sB}^{\otimes 2}  ) ) \nonumber \\
& = {\rm Tr}_{\sA} (\ketbra{\psi}{\psi}_\sA^{\otimes 2} ( \sum_i \alpha_i( S_\sA (\rho_{\sA| \phi}^{x_i} \otimes \rho_{\sA| \phi}^{x_i}) S_\sA))) \ .
\end{align}

By Lemma \ref{stateform} $( S_\sA (\rho_{\sA| \phi}^x \otimes \rho_{\sA| \phi}^x) S_\sA$ is of the form $\sum_j \beta_j \ketbra{\phi_j}{\phi_j}^{\otimes 2}$ with $\beta_j \geq 0$. Hence $ \hat F_\sA' = ( \sum_i \alpha_i( S_\sA (\rho_{\sA| \phi}^{x_i} \otimes \rho_{\sA| \phi}^{x_i}) S_\sA)$ is a valid effect on $\sA$ as long as its complement is also of the form $\sum_i \gamma_i \ketbra{\phi_i}{\phi_i}^{\otimes 2}$. 
Since the complement of $\hat F_{\sA \sB}$ is of the form $\sum_i \alpha_i \ketbra{x_i}{x_i}_{12} \otimes \ketbra{x_i}{x_i}_{34}$ it follows that the associated effect on $\sA$ (which is the complement of $\hat F_\sA'$) is of the required form. From this it follows that $\hat F_\sA'$ is a valid effect.
The set $\mathcal F_{d_\sA d_\sB}^\sg$ also contains effects $F_\sA \star F_\sB$ which are not necessarily of the form given above. However since these are product effects they trivially are consistent with {\bf C5}.

\end{document}